\RequirePackage[l2tabu]{nag} 
\documentclass[10pt]{article}

\usepackage{fullpage}
\usepackage{float}
\usepackage{fancyhdr} 
\usepackage{lastpage} 
\usepackage{extramarks} 
\usepackage{courier} 
\usepackage[top=1.3in, bottom=1.4in, left=1.4in, right=1.4in]{geometry}

\usepackage[utf8]{inputenc}
\usepackage[T1]{fontenc}

\usepackage{lmodern}
\usepackage[tracking=true,kerning=true,final]{microtype}
\usepackage{bbm}

\usepackage{bm}

\usepackage{graphicx} 
\usepackage{caption}
\usepackage{subcaption}
\usepackage[caption=false]{subfig}
\usepackage{enumitem}
\usepackage{soul,color}
\usepackage[utf8]{inputenc}
\usepackage[english]{babel}

\usepackage{amsmath,amsfonts,amsthm,amsbsy,amssymb}
\usepackage{mathtools}
\usepackage{bigints,bbold}

\DeclareMathOperator{\Tr}{Tr}

\begin{document}

\title{Reduced-space Gaussian Process Regression for Data-Driven Probabilistic Forecast of Chaotic Dynamical Systems}
\author{Zhong Yi Wan, Themistoklis P. Sapsis\medskip\thanks{Corresponding author: {sapsis@mit.edu},
Tel: (617) 324-7508, Fax: (617) 253-8689%
}\\
Department of Mechanical Engineering,
\\ Massachusetts Institute of Technology, \\
77 Massachusetts Ave., Cambridge, MA 02139}
\date{\today}
\maketitle

\begin{abstract}
We formulate a reduced-order strategy for efficiently forecasting complex high-dimensional dynamical systems entirely based on data streams. The first step of our method involves reconstructing the dynamics in a reduced-order subspace of choice using Gaussian Process Regression (GPR). GPR simultaneously allows for  reconstruction of the vector field and more importantly, estimation of local uncertainty. The latter is due to i) local interpolation error and ii) truncation of the high-dimensional phase space. This uncertainty component can be analytically quantified in terms of the GPR hyperparameters. In the second step we formulate stochastic models that explicitly take into account the reconstructed dynamics and their uncertainty. For regions of the attractor which are not sufficiently sampled for our GPR framework to be effective, an adaptive blended scheme is formulated to enforce correct statistical steady state properties, matching those of the real data. We examine the effectiveness of the proposed method to complex systems including the Lorenz 96, the Kuramoto-Sivashinsky, as well as a prototype climate model. We also study the performance of the proposed approach as the intrinsic dimensionality of the system attractor increases in highly turbulent regimes.  
\end{abstract}




\section{Introduction}

A broad range of systems are characterized by a high-dimensional phase space
and existence of persistent or intermittent instabilities. These
properties are ubiquitous in many complex systems involving fluid flows such
as the atmosphere, ocean, coupled climate system, confined
plasmas, and engineering turbulence at high Reynolds numbers. For these systems
short term prediction, as well as quantification of long-term statistics
can be very challenging. The difficulty is the result of i) intrinsic
limitations of typical order-reduction methods for systems exhibiting unstable
dynamics mitigated by strongly nonlinear energy transfers \cite{sapsis_majda_mqg,
sapsis_majda_tur, sapsis_majda_qgdo}, and ii) inevitable error in the
model equations especially when we do not have a complete
understanding of the underlying physical mechanisms \cite{majda11, Majda_filter,
branic_majda, Gershgorin201032}. 

For such systems it is often beneficial, instead of adopting the typical
equation-driven approach, to consider a data-driven perspective. One of the
simplest methods used to forecast dynamical systems is the vector autoregression
model \cite{kalman60, kalman61}, which assumes that the current state of the system depends linearly on a fixed number of its preceding states and
an additional noise term. These models have been applied to modeling problems in
economics, medicine and soil sciences \cite{kitagawa84, jones84, shumway88}
with success. Nevertheless, they are greatly limited by their \textit{ad hoc} parametric structure, which requires significant tuning and testing to be effectively applied to different systems. Another relevant direction involves the generation of symbolic nonlinear equations using time series of the system response \cite{Bongard2007, Schmidt09}. However, symbolic regression is expensive and to this end a sparse identification method has been recently proposed and demonstrated to work effectively for systems with low-dimensional attractors \cite{brunton16}. For turbulent systems with high-dimensional structures, physics-constrained nonlinear regression models have been developed in \cite{Majda2013} and \cite{Peavoy15} and shown to perform robustly. However, these models assume a stable linear part in the dynamics operator,  which is not always the case for the dynamics within a reduced order subspace \cite{sapsis_majda_mqgdo, sapsis_majda_mqg, sapsis_majda_tur}.

A class of non-parametric methods are inspired by an empirical forecasting
technique called analog forecasting. This approach was introduced in \cite{lorenz69}
as a method for predicting the time evolution of observables in dynamical
systems based on a historical record of training data. For any arbitrary
current state of the system, an analog, i.e. the state in the
historical record which most closely resembles the current state, is identified. Then, in
order to perform prediction, the historical evolution of the analog state is followed
for the desired lead time, and the observable of interest is predicted based
on the corresponding analog value. This method has been applied in a number of applications
with very good results \cite{zoth89, sugihara90, xavier07}.  However, its
main drawback is the difficulty to pick the most skillful analog that will
best represent the evolution of the system. This is because in analog
forecasting very little emphasis is placed on the geometrical structure of
the data-points.

Recent approaches \cite{Berry2015, Berry2015b, Berry2016} have successfully
managed to incorporate the geometrical properties of available data through use of \textit{diffusion maps} \cite{coifman05a, coifman06}. The basic idea is to represent the semigroup solution using a basis adapted to the invariant measure. Because of the `global' nature of the basis elements produced by the diffusion maps algorithm, a large number of basis elements is often necessary to probabilistically evolve initial states with small variance. In addition, computing these basis elements using diffusion maps is an expensive process especially for large data sets, as it requires solving an eigensystem whose size is proportional to that of the training data.

A different perspective for the same problem can be found in \cite{Sonday2010, Chiavazzo2014}. The idea is to utilize nonlinear diffusion map coordinates and formulate a deterministic dynamical system on the system manifold. This approach results in a reduced-space data-driven dynamical system which describes the evolution of states on the attractor. Models formulated this way are able to effectively take advantage of the low intrinsic dimensionality. Along the same spirit, in \cite{zhao15} the local dynamics on the attractor are represented with weighted averages of the data-points using local similarity kernels. However, the resulting low-dimensional attractors are typically noisy due to truncation errors, observation noise, or under-sampled training data. Therefore, it is crucial to be able to quantify the error in the evolved dynamics in reduced space. This is the main goal of this work.

In particular, we propose a generally applicable methodology for forecasting
and quantifying uncertainty in reduced-space states, based
on Gaussian process regression (GPR) \cite{rasmu05, forrester08,
perdikarisPRSA2015,bilionis12,PChen15}. The main advantage of employing GPR to reconstruct
the reduced-order dynamics is the simultaneous
estimation of the dynamics and the associated error/uncertainty,
which can be important when reduction dimension is lower than the intrinsic dimension of the system.
Using GPR dynamics we formulate a reduced-order stochastic model taking into account uncertainty from various sources. We also propose an adaptive blended scheme for systems which are not sufficiently sampled everywhere. We examine the effectiveness of the proposed method for different systems and evaluate its performance as the intrinsic dimension
of the system attractor increases in highly turbulent regimes.

This paper is structured as follows. In section 2, we give an overview for
Gaussian Process Regression technique. Section 3 describes the use of GPR to construct
stochastic models in reduced-order space and perform probabilistic forecast.
In the same section we present the decomposition of the error into different
components and quantify those in terms of the GPR hyperparameters. The proposed
methodologies are applied to three complex systems in section 4. Finally,
section 5 provides a summary and brief discussion of possible future directions.


\section{An Overview of Gaussian Process Regression}

In this section, we present an overview of Gaussian Process Regression \cite{rasmu05},
appropriately formulated for forecasting dynamical systems. GPR works
under the probabilistic regression framework, which takes as input a training data set $\mathcal{D} = \{(y_n,\mathbf{x}_n), n
= 1,...,N\}$ of $N$ pairs of vector input $\mathbf{x}_n \in \mathbb{R}^L$
and \textit{noisy} scalar output $y_n$, and constructs a model that
generalizes well to the distribution of the output at unseen input locations.
The noise in the output models uncertainty due to factors external to $\mathbf{x}$, such as truncation or
observation errors. Here we assume that noise is additive, zero-mean, stationary and normally distributed, such
that
\begin{equation}
        \label{eq:probreg}
        \begin{aligned}
                & y = f(\mathbf{x}) + \epsilon, & \epsilon \sim \mathcal{N}(0,\sigma^2_{\text{noise}}),
        \end{aligned}
\end{equation}
where $\sigma^2_{\text{noise}}$ is the variance of the noise.

The primary idea behind GPR is to use a Gaussian process (GP) to represent
$f$, referred to as \textit{latent variables}. The input $\mathbf{x}$ plays the role of indexing these latent variables such that any finite collection $\{f(\mathbf{x}_1),...,f(\mathbf{x}_k)\}$ with unique indices follow a consistent Gaussian distribution. In this way, we limit
ourselves to only looking at functions whose values correlate with each other
in a Gaussian manner. In Bayesian framework, this is equivalent to putting
a GP prior over functions. Due to the consistency requirement, we are able
to make inference on function values corresponding to unseen inputs conveniently using
a finite set of training data.

A major advantage for using the Gaussian prior assumption is that functions
can be conveniently specified by a mean function $m(\mathbf{x})$ and a covariance
function $k(\mathbf{x},\mathbf{x'})$:

\begin{equation} \begin{aligned}
        m(\mathbf{x}) &= \mathbb{E}[f(\mathbf{x})],  \\
        k(\mathbf{x},\mathbf{x}') &= \mathbb{E}[(f(\mathbf{x})-m(\mathbf{x}))(f(\mathbf{x}')-m(\mathbf{x}'))],
        \label{eq:meancov}
\end{aligned} \end{equation}
 where $\mathbb{E}[\cdot]$ denotes expectation. The form of the mean function is important only in unobserved region of the input space and usually set to zero. The properties of the process is then
entirely dictated by the covariance function, which is by definition symmetric and positive semi-definite
when evaluated at any pair of points in the input space. The covariance function
typically contains a number of free parameters called \textit{hyperparameters} which define the prior distribution
on $f(\textbf{x})$. The most commonly used is the squared exponential covariance function

\begin{equation}
        \label{eq:sqexp}
        k(\textbf{x},\textbf{x}') = \theta_1 \mbox{exp}\left(-\frac{||\mathbf{x}-\mathbf{x}'||^2}{2\theta_2}\right),
\end{equation}
where $||\cdot||$ is a norm defined on the input space. Note that this covariance function decays rapidly
when evaluated at increasingly distant pairs of input $\mathbf{x}$ and $\mathbf{x}'$,
indicating weak correlations between $f(\mathbf{x})$ and $f(\mathbf{x}')$.
$\theta_1$ is a hyperparameter specifying the maximum allowable covariance.
$\theta_2$ is a strictly positive hyperparameter defining rate of decay
in correlation as points become farther away from each other. Another hyperparameter
$\theta_3$, which is not expressed explicitly in (\ref{eq:sqexp}), is used
to represent the unknown variance $\sigma_{\text{noise}}^2$ of the i.i.d noise $\epsilon$ in (\ref{eq:probreg}).

The hyperparameters $\{\theta_1, \theta_2,\theta_{3}\}$ are grouped together
as a vector $\boldsymbol{\theta}$ treated as the realization of a random
vector $\boldsymbol{\Theta}$. The realization that is most coherent with
the data set is selected using training data and then used to make predictions.
Methods for estimating these parameters are described in section \ref{sec:GPRhyp}.
 \\

\subsection{Prediction with GPR}

Assuming that the hyperparameters are known, inference is easily made. Denoting
the vector of training latent variables by $\mathbf{f}$ and the vector of test latent variables
by $\mathbf{f^*}$, we have the following joint Gaussian distribution:

\begin{equation}
        p(\mathbf{f},\mathbf{f}^*) = \mathcal{N}\left(\mathbf{0},
        \begin{bmatrix}
                K_{\mathbf{f},\mathbf{f}} & K_{*,\mathbf{f}} \\
                K_{\mathbf{f},*} & K_{*,*}
        \end{bmatrix}
        \right).
        \label{eq:pff}
\end{equation}
$K$ is the symmetric covariance matrix whose $ij$th entry is the covariance
between the $i$th variable in the group denoted by the first subscript and
the $j$th variable in the group denoted by the second subscript ($*$ is used
in place of $\mathbf{f}^*$ for short), computed using covariance function
$k(\cdot,\cdot)$ in (\ref{eq:sqexp}) and corresponding hyperparameters. For convenience, we do
not explicitly write out the conditioning on $\boldsymbol{\theta}$ in (\ref{eq:pff})
and other conditional probability expressions that follow in this section.

The conditional probability for the training observations $\mathbf{y}$ can
then be incorporated to find the posterior distribution for $\mathbf{f^*}$.
Because of the noise assumption, we have
\begin{equation}
        \label{eq:ygivenf}
        p(\mathbf{y}|\mathbf{f}) = \mathcal{N}(\mathbf{f},\sigma^2_{\text{noise}}I).
\end{equation}
Using Bayes rule, the joint posterior can be written as
\begin{equation}
        \label{eq:jointpost}
        p(\mathbf{f},\mathbf{f^*}|\mathbf{y}) = \frac{p(\mathbf{f},\mathbf{f}^*)p(\mathbf{y}|\mathbf{f})}{p(\mathbf{y})},
\end{equation}
which can be marginalized to find $p(\mathbf{f^*}|\mathbf{y})$:

\begin{equation}
        \label{eq:fpost}
        p(\mathbf{f^*}|\mathbf{y}) = \int p(\mathbf{f},\mathbf{f^*}|\mathbf{y})
d\mathbf{f} = \frac{1}{p(\mathbf{y})}\int p(\mathbf{f},\mathbf{f}^*)p(\mathbf{y}|\mathbf{f})
\; d\mathbf{f}.
\end{equation}

This corresponds to conditioning the joint Gaussian prior distribution on
the observations, resulting in the closed-form Gaussian distribution

\begin{equation}
        \label{eq:fpostGaussian}
        p(\mathbf{f^*}|\mathbf{y}) = \mathcal{N}\big(K_{*,\mathbf{f}}(K_{\mathbf{f},\mathbf{f}}+\sigma^2_{\text{noise}}I)^{-1}\mathbf{y},K_{*,*}-K_{*,\mathbf{f}}(K_{\mathbf{f},\mathbf{f}}+\sigma^2_{\text{noise}}I)^{-1}K_{\mathbf{f},*}\big)
\triangleq \mathcal{N}(\overline{\mathbf{f}}^*,\Sigma_{\mathbf{f}^*}),
\end{equation}
for which a detailed derivation/proof
can be found in section 4.3.4 of \cite{murphy12}. Since $\mathbf{y^*} = \mathbf{f^*}
+ \boldsymbol{\epsilon^*}$, with $\boldsymbol{\epsilon}^* \sim \mathcal{N}(\mathbf{0},\sigma^2_{\text{noise}}I)$
being independent of $\mathbf{f}^*$, the mean and covariance can be directly
added to obtain

\begin{equation}
        \label{eq:ypostGaussian}
        p(\mathbf{y^*}|\mathbf{y}) = \mathcal{N}\big(K_{*,\mathbf{f}}(K_{\mathbf{f},\mathbf{f}}+\sigma^2_{\text{noise}}I)^{-1}\mathbf{y},K_{*,*}-K_{*,\mathbf{f}}(K_{\mathbf{f},\mathbf{f}}+\sigma^2_{\text{noise}}I)^{-1}K_{\mathbf{f},*}+\sigma^2_{\text{noise}}I\big)
\triangleq \mathcal{N}(\overline{\mathbf{f}}^*,\Sigma_{\mathbf{y}^*}).
\end{equation}

The computation complexity of (\ref{eq:ypostGaussian}) appears to be dominated
by the matrix inversion term $(K_{\mathbf{f},\mathbf{f}}+\sigma^2_{\text{noise}}I)^{-1}$.
However, if we use the same set of training cases, $K_{\mathbf{f},\mathbf{f}}$
remains the same and the inversion can be easily pre-computed in terms of the Cholesky factors and stored for all later uses. The overall complexity is then reduced to $\mathcal{O}(N^2)$. This makes
it feasible to use up to more than ten thousand of training examples to make
predictions. \\

\subsection{Evaluation of Hyperparameters} \label{sec:GPRhyp}

A crucial part of the GPR framework is choosing suitable hyperparameters
$\boldsymbol{\theta}$. The parameter choice has fundamental impact on how well the model fits with data. The best set of parameters are usually
obtained by optimizing over training data using appropriate objective/penalty
functions. In the following, two possible approaches are described. 

\subsubsection{Maximum a Posteriori Estimates}

A complete Bayesian approach involves placing a prior distribution (hyper-prior) $h(\boldsymbol{\theta})$ over the hyperparameters
and marginalize to eliminate the dependence of hyperparameters $\boldsymbol{\theta}$.
However, this process is usually computationally expensive. Instead, the
\textit{maximum a posteriori} (MAP) estimate is often used as a point estimate for
$\boldsymbol{\theta}$. If we assume uniform distributions on the hyperparameters,
the resulting MAP turns into a maximum likelihood estimate (MLE) for $\boldsymbol{\theta}$.
As a function of $\boldsymbol{\theta}$, the log-likelihood for the training
data can be written as

\begin{equation}
        \label{eq:loglik}
        \begin{aligned}
        \mathcal{L}_{\text{MAP}}(\boldsymbol{\theta}) = \log p(\mathbf{y}|\boldsymbol{\theta})
&= \log{\left[(2\pi)^{-\frac{k}{2}}|K_{\mathbf{y},\mathbf{y}}|^{-\frac{1}{2}}\exp{\left(-\frac{1}{2}\mathbf{y}^T\Sigma_{\mathbf{y}}\mathbf{y}\right)}\right]}\\
        &= \frac{1}{2}\textbf{y}^T (K_{\mathbf{y},\mathbf{y}})^{-1} \textbf{y}
- \frac{1}{2} \log |K_{\mathbf{y},\mathbf{y}}| - \frac{N}{2} \log 2\pi,
        \end{aligned}
\end{equation}
where $N$ is the size of the training data and $K_{\mathbf{y},\mathbf{y}}
\triangleq K_{\mathbf{f},\mathbf{f}}+\sigma_{\text{noise}}I$. The partial
derivatives with respect to the hyperparameters can be readily obtained by
differentiating (\ref{eq:loglik}) and simplifying with relevant matrix identities

\begin{equation}
        \label{eq:loglik_partial}
        \begin{aligned}
                \frac{\partial \mathcal{L}_{\text{MAP}}}{\partial \theta_j}
&= \frac{1}{2}\mathbf{y}^T(K_{\mathbf{y},\mathbf{y}})^{-1}\frac{\partial
K_{\mathbf{y},\mathbf{y}}}{\partial \theta_j}K_{\mathbf{y},\mathbf{y}}\mathbf{y}-\frac{1}{2}\Tr{\left((K_{\mathbf{y},\mathbf{y}})^{-1}\frac{\partial
K_{\mathbf{y},\mathbf{y}}}{\partial \theta_j}\right)} \\
                &= \frac{1}{2}\Tr{\left((\boldsymbol{\alpha}\boldsymbol{\alpha}^T-(K_{\mathbf{y},\mathbf{y}})^{-1})\frac{\partial
K_{\mathbf{y},\mathbf{y}}}{\partial \theta_j}\right)},
        \end{aligned}
\end{equation}
where $\boldsymbol{\alpha} \triangleq (K_{\mathbf{y},\mathbf{y}})^{-1}\mathbf{y}$.
The most computationally expensive step in this expression is evaluating
$(K_{\mathbf{y},\mathbf{y}})^{-1}$ which requires $\mathcal{O}(N^3)$ time,
but again only needs to be computed once for all $\theta_j$. Thus the overall time requirement
for computing the analytical derivative of the log likelihood is only $\mathcal{O}(N^2)$ per hyperparameter,
making it realistic to implement a gradient-based optimization algorithm. For the examples in this work we implement a conjugate gradient (CG) optimizer to perform searches for the optimal hyperparameters.\\

\subsubsection{Cross-validation}

An alternative approach that emphasizes more on empirical performance for the selection of hyperparameters is the cross-validation
method. The whole data set is split into a training set and a validation
set, and the prediction performance of GPR models built with the training
set is measured on the validation set. The process is usually repeated for
a large number of different partitions of training and validation data in
order to obtain an unbiased measure overall. The special case of using one
data point for validation and all others for training is called Leave-one-out
(LOO) cross-validation. In LOO, the predictive log-likelihood for a single
validation case $y_n$ is

\begin{equation}
        \label{eq:LOOloglik}
        \log{p(y_n|\mathbf{y}_{-n},\boldsymbol{\theta})} = -\frac{1}{2}\log{\sigma_{n}^2}-\frac{y_n-\mu_n}{2\sigma_{n}^2}-\frac{1}{2}\log{2\pi},
\end{equation}
where $\mathbf{y}_{-n}$ denotes training data with case $n$ excluded. $\mu_n$
and $\sigma_n$ are predicted mean and variance for case $n$, computed using
(\ref{eq:ypostGaussian}), and can be considered functions of $\boldsymbol{\theta}$.
Summing over all cases, the total likelihood is then

\begin{equation}
        \label{eq:totLOOloglik}
        \mathcal{L}_{\text{LOO}}({\boldsymbol{\theta}}) = \sum\limits_{n=1}^{N}
\log{p(y_n|\mathbf{y}_{-n},\boldsymbol{\theta})},
\end{equation}
which can be optimized over $\boldsymbol{\theta}$. This expression may seem
expensive to compute at first because calculating each $\mu_n$ and $\sigma_n$
requires inverting a different matrix. However, these matrices are highly
similar, each with one row and column removed from the covariance matrix
for the entire training set. As a result, $\mu_n$ and $\sigma_n$ can be conveniently
computed from $(K_{\mathbf{y},\mathbf{y}})^{-1}$ as
\begin{equation}
        \label{eq:LOOmusigma}
        \begin{aligned}
                \mu_n &= y_n - \left[(K_{\mathbf{y},\mathbf{y}})^{-1}\mathbf{y}\right]_n/\left[(K_{\mathbf{y},\mathbf{y}})^{-1}\right]_{nn},
\\
                \sigma^2_n &= 1/\left[(K_{\mathbf{y},\mathbf{y}})^{-1}\right]_{nn},
        \end{aligned}
\end{equation}
where the subscripts refer to the indices of the corresponding matrix/vector.
Carefully examining (\ref{eq:LOOmusigma}) would reveal that $\mu_n$ is rightfully
independent of $y_n$. Since only one matrix inversion is required, the overall
complexity for calculating $\mathcal{L}_{\text{LOO}}$ is $\mathcal{O}(N^2)$.
\\

Taking the derivative of the expressions in (\ref{eq:LOOmusigma}) with respect
to $\theta_j$ we obtain

\begin{equation}
        \label{eq:LOOmusigma_d}
        \begin{aligned}
                \frac{\partial \mu_n}{\partial \theta_j} &= \frac{\left[Z_j\boldsymbol{\alpha}\right]_n}{\left[(K_{\mathbf{y},\mathbf{y}})^{-1}\right]_{nn}}
- \frac{\boldsymbol{\alpha}\left[Z_j(K_{\mathbf{y},\mathbf{y}})^{-1}\right]_{nn}}{\left[(K_{\mathbf{y},\mathbf{y}})^{-1}\right]_{nn}},
\\
                \frac{\partial \sigma^2_n}{\partial \theta_j} &= \frac{\left[Z_j(K_{\mathbf{y},\mathbf{y}})^{-1}\right]_{nn}}{\left[(K_{\mathbf{y},\mathbf{y}})^{-1}\right]^2_{nn}},
        \end{aligned}
\end{equation}
where $Z_j \triangleq K_{\mathbf{y},\mathbf{y}}^{-1}\frac{\partial K_{\mathbf{y},\mathbf{y}}}{\partial
\theta_j}$ and $\boldsymbol{\alpha}$ is defined as in previous section. Using
the chain rule, the derivative of the LOO likelihood is then
\begin{equation}
        \label{eq:LOOlik_d}
        \begin{aligned}
                \frac{\partial \mathcal{L}}{\partial \theta_j} &= \sum\limits^N_{n=1}\frac{\partial
\log{p(y_n|\mathbf{y}_{-n},\boldsymbol{\theta})}}{\partial \mu_n}\frac{\partial
\mu_n}{\partial \theta_j}+\frac{\partial \log{p(y_n|\mathbf{y}_{-n},\boldsymbol{\theta})}}{\partial
\sigma^2_n}\frac{\partial \sigma^2_n}{\partial \theta_j} \\
                &= \sum\limits^N_{n=1} \left(\alpha_i[Z_j\boldsymbol{\alpha}]_n-\frac{1}{2}\left(1+\frac{\alpha_n^2}{\left[(K_{\mathbf{y},\mathbf{y}}^{-1}\right)]_{nn}}\right)\left[Z_j(K_{\mathbf{y},\mathbf{y}})^{-1}\right]_{nn}\right)/\left[(K_{\mathbf{y},\mathbf{y}})^{-1}\right]_{nn}.
        \end{aligned}
\end{equation}
The computational complexity is $\mathcal{O}(N^3)$, dominated by the $N \times
N$ matrix multiplication calculating $Z_j$. Thus, using a gradient based
optimization algorithm for LOO cross-validation is more expensive than the
MAP estimate. However, due to the fact that it is formulated to minimize
the prediction error, LOO cross-validation generally has better performance
empirically. For this reason, we can first use MAP to coarsely find the hyperparameters
and fine-tune them using LOO cross-validation gradients. The combination
of both approaches ensures that good performances are attained at moderate
computational costs.\\


\section{Reduced-order data-driven forecast models}

Building on the GPR framework, we propose a purely data-driven method for
constructing reduced-order dynamical models for nonlinear chaotic systems.
It is assumed that we have no access to the analytical expressions for the
vector field representing the dynamics of states; instead, we only have access
to some sample data consisting of the state vector $\mathbf{u}$
and its rate of change $\dot{\mathbf{u}}$: $\mathcal{D} = \{\mathbf{u}^{(n)},\dot{\mathbf{u}}^{(n)},
n = 1,...,N \;|\; \mathbf{u}, \dot{\mathbf{u}} \in \mathbb{R}^D\}$. We also
assume that the data originate from an ergodic system $\dot{\mathbf{u}} = g(\mathbf{u})$
that has reached its statistical steady state. In addition we assume, for the
purpose of algorithm training that the data points and are noise-free. It
is not required that $\mathbf{u}$ and $\dot{\mathbf{u}}$ are arranged in time
order. 

We are primarily interested in the case where the states $\mathbf{u}_i$
are in general high-dimensional but `live' on a \textit{manifold} with low intrinsic
dimensionality. However, we also assess the performance of the developed framework when this is not necessarily the case. To construct a predictive dynamical model from $\mathcal{D}$ satisfying
these assumptions we follow three major steps:

\begin{enumerate}
        \item Derive reduced-order representations/embeddings $\mathbf{y}
\in \mathbb{R}^d$ and $\dot{\mathbf{y}} \in \mathbb{R}^d$ of the given data
 $\mathbf{u} \in \mathbb{R}^D$ and $\dot{\mathbf{u}}\in \mathbb{R}^D$, where
$d \ll D$.
        \item Learn GPR models for \textit{each} component of $\dot{\mathbf{y}}$
as a function of the reduced state $\mathbf{y}$.
This will result in GPR models with independent hyperparameters for each component $\dot y_i, i=1,...,d$
in the reduced-order space.        \item Formulate stochastic models for
$\mathbf{y}$ in the reduced-order space using GPR dynamics.
\end{enumerate}

\subsection*{Step 1: Order-reduction}

The first step aims to provide a mapping from each state vector $\mathbf{u}^{(n)}$
in the $D$-dimensional ambient space to a representation $\mathbf{y}$ in
a reduced $d$-dimensional space. In doing so, the intrinsically low dimensional
structure of the data is extracted and exploited to facilitate efficient
modeling. The procedure for finding such a map is generally known as \textit{dimensionality
reduction} and is by itself an active area of research. Many techniques, linear and nonlinear,
have been established and used with great success in a variety of applications.
It is not our focus in this work to select the best possible procedure for each application. Instead, we will only use a few state-of-the-art methods and assess their performance when integrated with the GPR data model.

After dimension reduction, we obtain the low-dimensional coordinates for all training data, and more importantly
a mapping $\psi(\mathbf{u})$ \footnote{can have no explicit function form and instead
depends on training data, which are not included in the function argument
for clarity} which can be used to convert any non-training data to its low-dimensional
representation. We can then easily use this map along with ambient space
dynamics $\dot{\mathbf{u}}$ to find the reduced space dynamics by evaluating
the limit

\begin{equation}
        \dot{\mathbf{y}} = \lim_{\Delta t \to 0} \frac{\psi(\mathbf{u}+\Delta
t\dot{\mathbf{u}})-\psi(\mathbf{u})}{\Delta t}.
        \label{eq:ydot}
\end{equation}

\begin{figure}[t]
        \centering
        \begin{subfigure}[h]{.49\textwidth}
                \includegraphics[width=\linewidth]{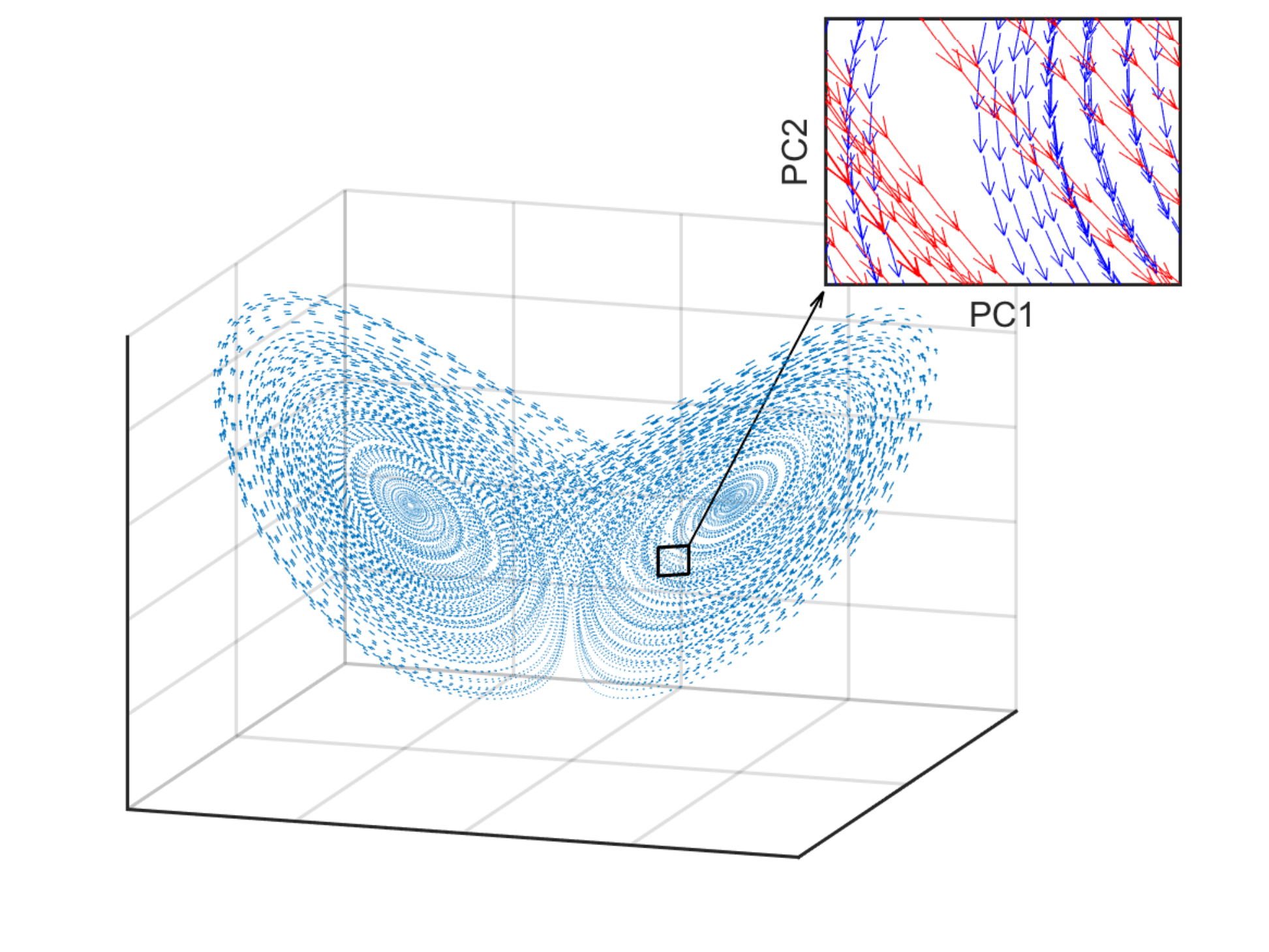}
          \end{subfigure}
          \begin{subfigure}[h]{.49\textwidth}
                \includegraphics[width=\linewidth]{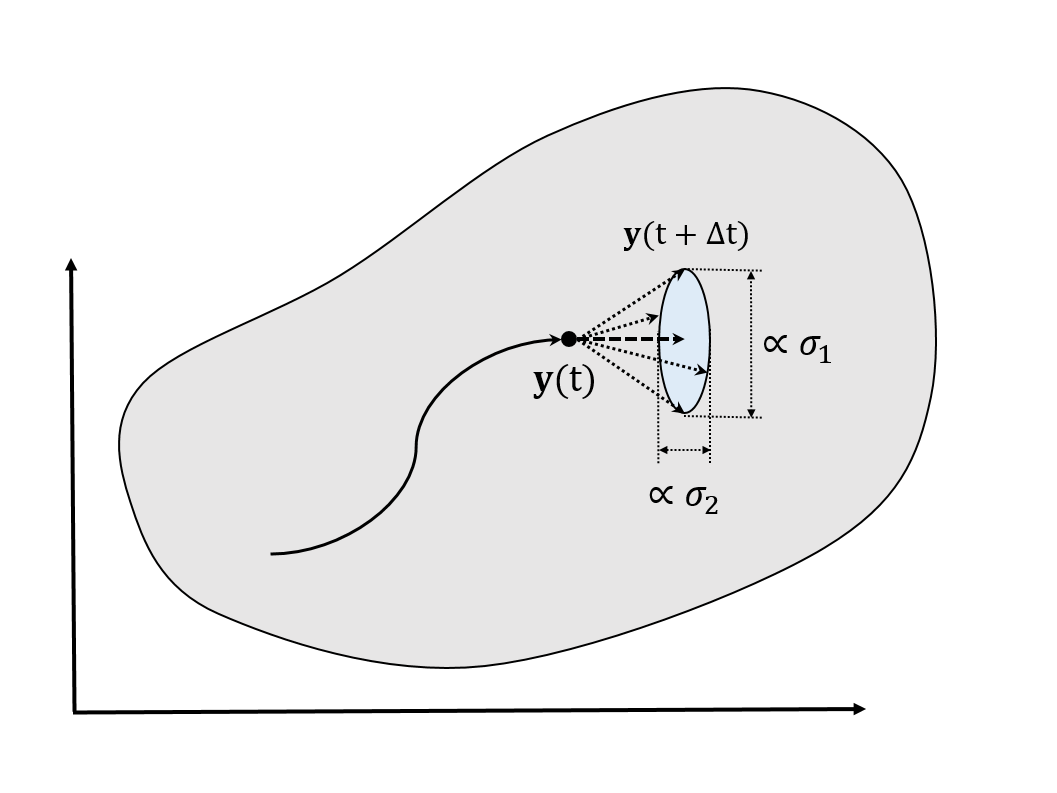}
          \end{subfigure}
          \caption{[left] Lorenz 63 attractor: in reduced state space, uncertainty in the reconstructed
vector field is observed; zoomed-in plot shows two groups of intersecting/conflicting trajectories (red and blue) in 2-dimensional PCA projected space; without additional information, a given state has two equally likely future states in the projected space [right] probabilistic modeling
of dynamics in a 2-dimensional reduced space: a point is marched to a Gaussian
distribution; mean is connected with long-dashed arrow; aspect ratio of equiprobability
contours is characterized by the variances predicted by the GPR.}
          \label{fig:nonuniquedynamics}
\end{figure}

\medskip
Here we have assumed that dimension reduction is performed
`imperfectly', as often is the case in realistic applications.  Through dimension
reduction, the state vectors are usually transformed in a way that
allows us to rank the coordinates by their importance and truncate the ones
deemed `unimportant' by the ranking criteria. However, it is realistically
impossible for the dynamics to be decoupled during the same process. As a
result, the truncated coordinates still play roles in deciding the dynamics
of the preserved coordinates, which leads to the unfortunate existence of
non-unique dynamics. Figure \ref{fig:nonuniquedynamics}
(left) illustrates this phenomenon, using the Lorenz 63 system. The attractor
has a fractal dimension that is slightly higher than 2 and the principal
component analysis is capable of generating a two-dimensional embedding containing 96\% of the
total variance. Plotting the trajectories in the reduced space reveals the presence of non-unique
dynamics as suggested by intersecting trajectories. Furthermore, due to the fact that dynamics
are constructed with a finite number of training examples, predictions are
made with different levels of confidence depending on the location of the input
with respect to the training data (i.e. potential interpolation error). These two factors
motivate the use of a probabilistic description for the reduced-order dynamics.
This makes a GPR-based dynamics model particularly suitable in reduced-order
space.

\subsection*{Step 2: Gaussian Process Regression}

In the second step, we train $d$ independent GPR models, one for each component
of the reduced dynamics $\dot{\mathbf{y}}$, by estimating the corresponding
hyperparameters with the methods described in section \ref{sec:GPRhyp}. The
models take in all components of the reduced state vector $\mathbf{y}$ as
the common input. We can write the GPR models in the form

\begin{equation}
        \dot{y}_i = \mathcal{G}_i(\mathbf{y},\mathcal{D}_i,\boldsymbol{\theta}_i)
= \mathcal{N}\bigg(f_i(\mathbf{y},\mathcal{D}_i,\boldsymbol{\theta}_i),\sigma^2_i(\mathbf{y},\mathcal{D}_i,\boldsymbol{\theta}_i)\bigg),
        \label{eq:GPRdynamics}
\end{equation}
where $\mathcal{D}_i$ denotes the training data set $\{\dot{y}_i^{(n)},
\mathbf{y}^{(n)}| n = 1,...,N\}$. In the theorem below we will show that
such a model offers a clean way to quantify the uncertainties in the predicted
dynamics arising from both of the major sources, namely interpolation
and truncation errors.\\

\textbf{Theorem 1.} \textit{For the GPR dynamical model (\ref{eq:GPRdynamics})
with a squared exponential covariance function of the form (\ref{eq:sqexp}),
the variance of the estimated derivative is bounded by the observation noise
and the first hyperparameter as follows:}
\begin{equation*}
\sigma^2_i(\mathbf{y},\mathcal{D}_i,\boldsymbol{\theta}_i) \leq \theta_{i,1}+\theta_{i,3},
\end{equation*}
\textit{where $\theta_{i,1}$ denotes the hyperparameter $\theta_1$ for model
$\mathcal{G}_i$ and $\theta_{i,3}=\sigma^2_{\text{noise},i}$. The first term
on the right hand side accounts for uncertainties
arising from interpolating discrete training data, while the second component
accounts for uncertainty due to order-reduction and/or observation errors.}\\

\textbf{Proof:} Let us consider the predicted variance for a single test
case using (\ref{eq:ypostGaussian})

\begin{displaymath}
        (\sigma_i^*)^2 = k(\mathbf{y^*},\mathbf{y^*})-\mathbf{k}_{*,\mathbf{f}}(K_{\mathbf{f},\mathbf{f}}+\sigma_{\text{noise},i}^2I)^{-1}\mathbf{k}_{\mathbf{f},*},
\end{displaymath}
where $k(\mathbf{y^*},\mathbf{y^*}) = \theta_{i,1}+\theta_{i,3}$ and $\mathbf{k}_{*,\mathbf{f}}$,
${\mathbf{k}_{\mathbf{f},*}}$ are row and column vectors respectively. Since
$k(\cdot,\cdot) \geq 0$, $\mathbf{k}_{*,\mathbf{f}} \geq \mathbf{0}$ and ${\mathbf{k}_{\mathbf{f},*}}
\geq \mathbf{0}$. Thus $(\sigma_i^*)^2$ is maximized when $\mathbf{k}_{*,\mathbf{f}}$,
${\mathbf{k}_{\mathbf{f},*}}$ approach $\mathbf{0}$. This corresponds to
the case where the test input $\mathbf{y}^*$ is distant from all training
inputs so that the exponents in the covariance function all approach $- \infty$.
The resulting $(\sigma_i^*)^2$ then takes the value $\theta_{i,1} + \theta_{i,3}$.

To understand the role of each term in the derived bound, we look at a hypothetical
`perfect' training data set for $\mathbf{y}^*$, where all training pairs
$(\mathbf{y}_{i},\dot {y}_i)$ have the same input as the test case,
i.e. $\mathbf{y} = \mathbf{y}^*$ for all $\mathbf{y} \in \mathcal{D}_i$.

Then, 
\begin{displaymath}
k = \theta_{i,1} \text{ for every entry in } \mathbf{k}_{*,\mathbf{f}}, \mathbf{k}_{\mathbf{f},*}
\text{ and } K_{\mathbf{f},\mathbf{f}}.
\end{displaymath}
As a result, $K_{\mathbf{f},\mathbf{f}}+\sigma^2_{\text{noise},i}I$ and its
inverse both have simple structures and it can be easily verified that 
\begin{displaymath}
\mathbf{k}_{*,\mathbf{f}}(K_{\mathbf{f},\mathbf{f}}+\sigma_{\text{noise},i}^2I)^{-1}\mathbf{k}_{\mathbf{f},*}
= \frac{N\theta_{i,1}^2}{\sigma^2+N\theta_{i,1}},
\end{displaymath}
where $N$ is the number of training cases and the dimension of $K_{\mathbf{f},\mathbf{f}}$.
Thus we have the limit\begin{displaymath}
\lim_{N \to \infty}\mathbf{k}_{*,\mathbf{f}}(K_{\mathbf{f},\mathbf{f}}+\sigma_{\text{noise},i}^2I)^{-1}\mathbf{k}_{\mathbf{f},*}
= \theta_{i,1}.
\end{displaymath} 
It follows that 
\begin{displaymath}
\lim_{N \to \infty}(\sigma_i^*)^2 = \theta_{i,1}-\theta_{i,1}+\theta_{i,3}=\theta_{i,3}.
\end{displaymath}
This corresponds to the minimum value of $\sigma_i^{*2}$, where the interpolation
process (GPR) causes zero uncertainty in the predicted dynamics. The entire
variance can be attributed to external factors, primarily  the effect
of the truncated coordinates in this case. This completes the proof. 

\subsection*{Step 3: Formulation of the stochastic model}

We can now use GPR dynamics to construct stochastic models. The simplest and most natural approach is to utilize a diffusion process with drift and diffusion coefficients obtained directly
from the GPR:

\begin{equation}
        \left(dy_i\right)_{\text{GPR}} = f_i(\mathbf{y},\mathcal{D}_i,\boldsymbol{\theta}_i)
dt + \sigma_i(\mathbf{y},\mathcal{D}_i,\boldsymbol{\theta}_i) dW_i,
        \label{eq:model3}
\end{equation}
where $W_i$ denotes a standard Wiener process. The drift and diffusion coefficients
are the mean and uncertainty of the GPR models $\mathcal{G}_i$. By numerically solving
(\ref{eq:model3}) for an ensemble of Monte Carlo samples, the reduced-order
states can be forecasted along with associated uncertainty quantifications.
Note that more complicated stochastic models may be utilized in order to take into
account time correlation of the noise. However, in the present context
we resort to the most straightforward option of uncorrelated noise with spatially
non-homogenous intensity.


\subsection{Benchmark: Mean stochastic model (MSM) }

We now give a brief overview of the mean stochastic models approach. This
is simple but yet powerful method for uncertainty quantification and filtering
 of reduced-order set of variables describing turbulent systems  \cite{Majda_filter}.
The method, although data-driven, is very different in nature from the presented
framework since it relies on the global statistics
of the system attractor, rather than local information. Nevertheless, a
comparison with the presented approach will reveal the types of systems and dynamical
regimes where it is advantageous to use local dynamical information through
GPR instead of the much cheaper MSM method.   

In MSM the main idea is to build up the model with two statistical equilibrium
properties, the energy spectrum and damping time scales for each considered
variable. In particular, the components of the reduced state are modeled
as diffusion processes having the form 

\begin{equation}
        \left(dy_i\right)_{\text{MSM}} = c_i y_i dt + \xi_i dW_i,
        \label{eq:msm3}
\end{equation}where $c_i$ and $\xi_i$ are constants calculated from data,
making (\ref{eq:msm3}) an Ornstein-Uhlenbeck (OU) process. In the statistical
steady state, model variable $y_i$ has zero mean\footnote{which we assume
to match that of the real data, as satisfied automatically by many dimension
reduction methods; in case this is not true for the reduced data, it will
need centering and the modeled quantity instead measures the deviation
from the real data mean}. Its energy/variance $E_i$ and decorrelation time
scaled $T_i$ are given by

\begin{equation}
        \begin{aligned}
                E_{i,\text{MSM}} &= \text{var}\,(y_i)_{\text{MSM}} = -\xi_i^2/(2c_i),
& T_{i,\text{MSM}} = -1/c_i.
        \end{aligned}
        \label{eq:OUvardt}
\end{equation}On the other hand, energy and decorrelation time of the data
set can be calculated by

\begin{equation}
        \begin{aligned}
                E_i &= \text{var}\,(y_i) = \mathbb{E} \left[y_i\overline{y_i}
\right] \ \text{ and } \
                T_i &= \int\limits_0^{\infty} \frac{\mathbb{E}[y_i(t)\overline{y_i(t+\tau)}]}{\text{var}\,(y_i)}
\, d\tau,
        \end{aligned}
        \label{eq:MSMvardt}
\end{equation}where the overhead bar represents taking the complex conjugate
if the reduced modes are complex-valued. Assuming this information is readily
available, the variance and decorrelation time of the MSM can be matched
with those of the real data by letting
\cite{Majda_filter}
\begin{equation}
        \begin{aligned}
                c_i &= -1/T_i, & \ \ \ \ \xi &= \sqrt{2E_i/T_i}. 
        \end{aligned}
        \label{eq:MSMconst}
\end{equation}By design, MSM behaves similarly to the real system in the
statistical steady state. The damping and diffusion components mimic the
nonlinear interactions between different modes by removing and injecting
energy into the system, respectively. The MSM can be especially effective
for highly chaotic systems because their behaviors are dominated by turbulent
nonlinear interactions whose energy flow bears greater resemblance to the
random process used in the model.

\subsection{Blended GPR-MSM forecast models}

One apparent drawback of the reduced-order GPR method is the absence of any
mechanism to ensure that trajectories stay on the  attractor. Indeed, Theorem 1 implies that GPR naturally
associates more uncertainty in `off-attractor' regions where no training data is present. Therefore, once a trajectory is driven into
such regions, close-to-zero drift coefficients and large diffusion coefficients will be repeatedly generated and
very likely will drive the trajectory farther away from the attractor. 

To address this issue, we  formulate a blended GPR - MSM approach: GPR model
is used in regions well encapsulated by training data or else MSM dynamics
is used. Note that the variance produced by model (\ref{eq:GPRdynamics})
offers an easy way to determine whether an excessive amount of uncertainty
is due to the training data not covering the current state very well, by
comparing it with the maximum possible interpolation uncertainty $\theta_{i,1}$.

Hence, we define the following composite indicator function to facilitate
selection of the more appropriate dynamical model:

\begin{equation}
        \chi(\mathbf{y}) = \prod\limits_{i=1}^d \mathbb{1}\left(\frac{\sigma^2_{i}(\mathbf{y})-\sigma^2_{\text{noise},i}}{\theta_{i,1}}
<\delta\right)
        \label{eq:indicator}
\end{equation}where $\delta$ is a threshold level between 0 and 1, and $\mathbb{1}(\cdot)$
is an indicator function which takes an expression as its input and returns
1 if the expression is true and 0 otherwise. 

\begin{figure}[t]
        \centering
        \includegraphics[width = .8\linewidth]{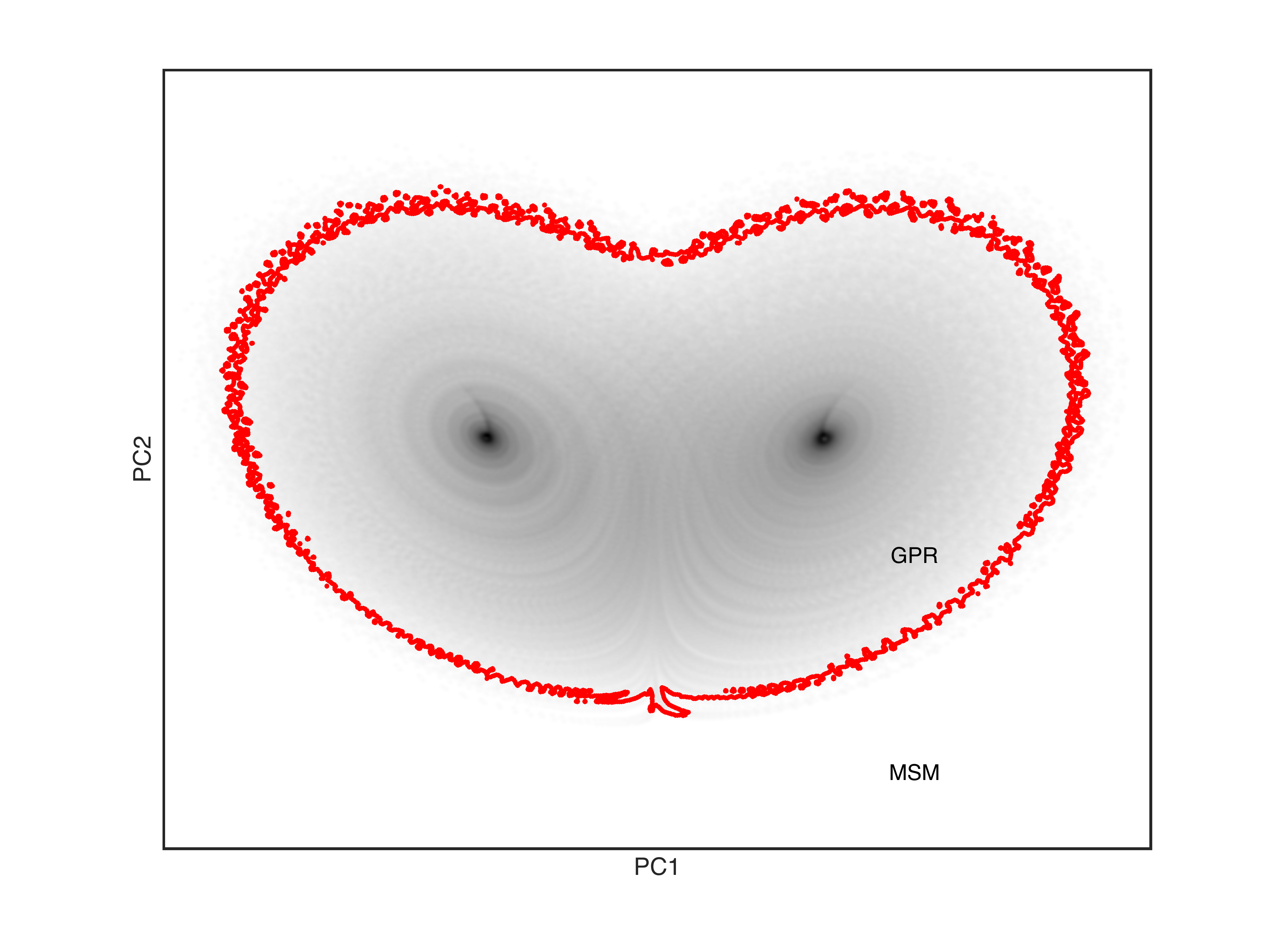}
        \caption{Blended GPR-MSM forecast for Lorenz 63: red dots are test
points whose
$\chi$ value (relative to a training data time series of size 10,000) are
close to the threshold $\delta = 0.2$ and they form a decision boundary in
the two-dimensional
reduced space. Dynamics are determined by GPR when a point lies inside the
boundary and by MSM otherwise. The corresponding two-dimensional joint density
for the training data, generated by a bivariate kernel estimator, is
shown
in the back.}
        \label{fig:mixedL63}
\end{figure}

In this way, $\chi$ returns a value of 1 only if the interpolation uncertainty
$\mathbf{y}$ falls below a certain percentage of its maximum possible value.
This function is equivalent to drawing a $d-1$ dimensional contour object in the reduced state space and
assigning all points inside to have function values of 1. This is a natural
estimate of the `trust region' in the reduced space where GPR dynamics is
supported by well positioned training data to produce reliable results. Figure
\ref{fig:mixedL63} shows an example of trust region in the
case of Lorenz 63 system. The resulting decision boundary is an `envelope'
that encloses the majority of the training data. In general, GPR dynamics
is used when interpolation with the available data is feasible and MSM is
used when extrapolation is required. Most importantly, the $\chi$ criterion
(\ref{eq:indicator}) is extremely cheap to evaluate and with almost no additional
computational expense besides computing the predicted variance. 

When implementing the forecast, GPR model is used if
$\chi$ evaluates to 1 on the current state and MSM model is used otherwise.  Following this
criterion, our blended-dynamics model can be written as

\begin{equation}
        \left(\frac{dy_i}{dt}\right)_{\text{mixed}} = \chi(\mathbf{y})\left(\frac{dy_i}{dt}\right)_{\text{GPR}}+(1-\chi(\mathbf{y}))\left(\frac{dy_i}{dt}\right)_{\text{MSM}}.
        \label{eq:mixedmodel}
\end{equation}
This  blended-dynamics model has the advantages of both GPR
and MSM dynamics. In short times scales dynamics are dominated by GPR to
produce more accurate short-term forecast, whereas for longer time scales
MSM dynamics dominate to ensure that key statistical properties of the ensemble match
with those of the real attractor. With this set of properties we have a data-driven scheme that is not necessarily characterized by a stable linear operator (as it is the case for example in physics-constrained models \cite{Majda2013, Peavoy15}), but on the other hand ensures stability of second-order statistics in the long time regime. This is particularly important for the dynamics within the reduced-order subspace of a turbulent system characterized by unstable linear dynamics (see e.g. \cite{sapsis_majda_mqgdo}).


\section{Applications}

In this section, we demonstrate the performance of our proposed methods in
three different applications originating from different fields of study and
exhibiting varying levels of chaotic behaviors. 

\subsection{Kuramoto-Sivashinsky equation}

We first study the Kuramoto-Sivashinsky (K-S) equation, originally developed
by Kuramoto to model the angular-phase turbulence of a reaction diffusion
system \cite{Kuramoto1976}, and by Sivashinsky to model perturbations of
a plane flame front propagating in a fuel-oxygen mixture \cite{Sivashinsky1977}.
Here we work with the one-dimensional K-S equation in \textit{derivative} form: 

\begin{equation} \begin{aligned}
        & \frac{\partial u}{\partial t} = -\nu\frac{\partial^4 u}{\partial
x^4} -\frac{\partial^2 u}{\partial x^2}-u\frac{\partial u}{\partial x}, \\
        & u(0,t) = u(L,t) = \frac{\partial u}{\partial x} \bigg |_{x=0} =
\frac{\partial u}{\partial x} \bigg |_{x=L} = 0, \\
        & u(x,0) = u_0(x),
        \label{eq:KSeq1}
\end{aligned} \end{equation}
where $u$ is the modeled quantity depending on spatial variable $x \in [0,L]$ and time variable $t \in [0,\infty)$. $\nu > 0$ is a physical
constant representing viscosity. Dirichlet and Neumann boundary conditions
are assigned to ensure that the system is ergodic \cite{blonigan14}. \\

(\ref{eq:KSeq1}) is discretized spatially with 2nd order
accurate finite difference schemes to produce a coupled system of ordinary
differential equations:

\begin{equation}
        \frac{du_i}{dt} = -\nu\frac{u_{i-2}-4u_{i-1}+6u_i-4u_{i+1}+u_{i+2}}{\Delta
x^4} - \frac{u_{i+1}-2u_{i}-u_{i-1}}{\Delta x^2}-\frac{u_{i+1}^2-u_{i-1}^2}{2\Delta
x},
        \label{eq:KSFD}
\end{equation}where $u_i$ represents the value of $u$ at the $i$th node,
i.e. $u_i = u(x_i) = u(i\Delta x) = u(\frac{iL}{D})$, $i = 0,1,...,D+1$ and
$D$ is the number of discretized fields in $[0,L]$. In this example we use $D = 512$. Boundary conditions
are satisfied by letting $u_{0} = u_{D+1} = 0$ and including additional ghost
nodes $u_{-1} = u_{1}$, $u_{D+2} = u_{D}$ to account for the Neumann boundary
conditions.

Data is simulated by solving (\ref{eq:KSFD}) for 5000 time units at intervals of 0.2, after an initial spin-up
period of 1000 units. The first 60\% of the data (15000 points) is used for
training, and the remainder is used for testing the prediction skills of
the proposed methods. 
\begin{figure}[t]
        \centering
          \includegraphics[width=\linewidth]{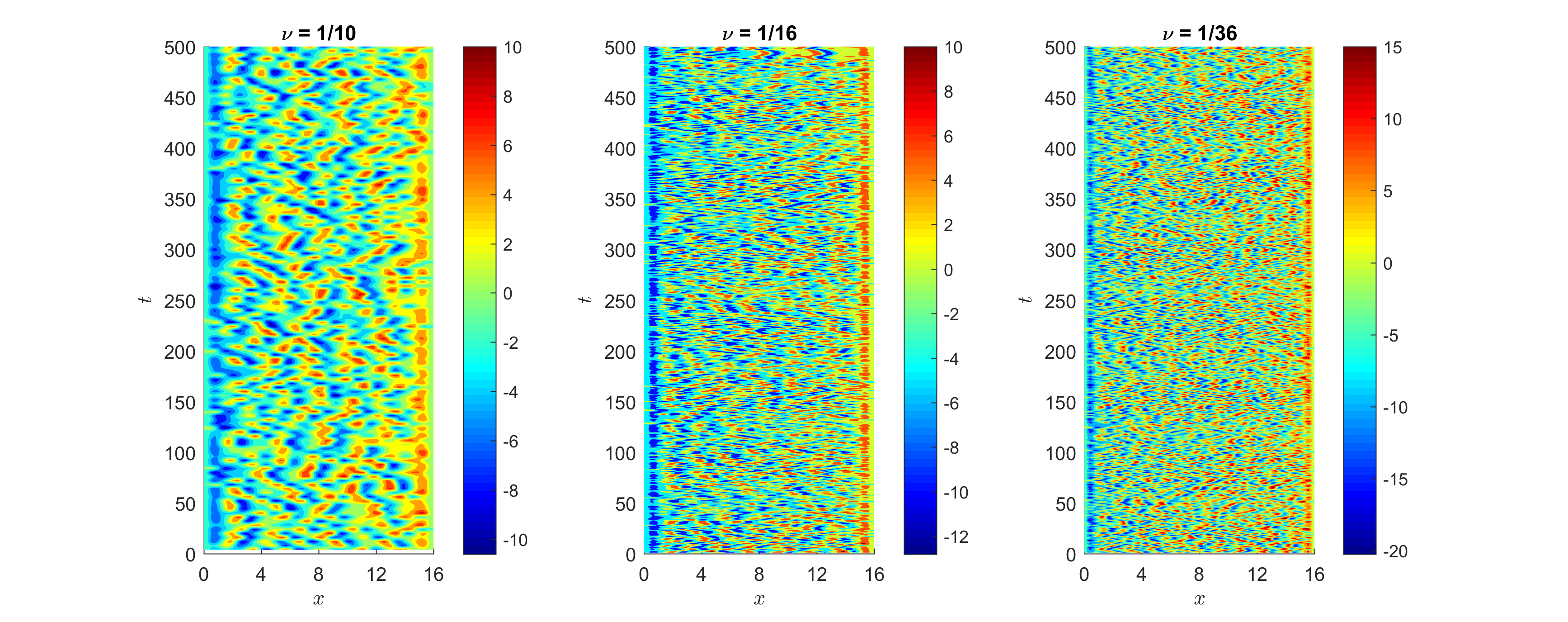}
          \caption{Contour plots for $u(x,t)$ at different $\nu$ values in steady state. The system becomes more chaotic for smaller values of $\nu$.}
          \label{fig:KScontour}
\end{figure}

The chaotic behavior of the K-S equation is dependent on the bifurcation
parameter $\tilde{L} = L/2\pi\sqrt{\nu}$ \cite{Kevrekidis90}. The system
displays higher levels of chaos for bigger values of $\tilde{L}$ (see Figure
\ref{fig:KScontour}). To test out the performance of our proposed methodology
in different chaotic regimes, $L$ is kept fixed at 16 and $\nu$ is changed
to produce different values of $\tilde{L}$ and thus differently chaotic systems.
Here we study the prediction skills for two particular systems: $\nu = 1/10$
and $\nu = 1/16$.

\begin{figure}[t]
        \centering
        \begin{subfigure}[h]{.7\textwidth}
                \includegraphics[width=\linewidth]{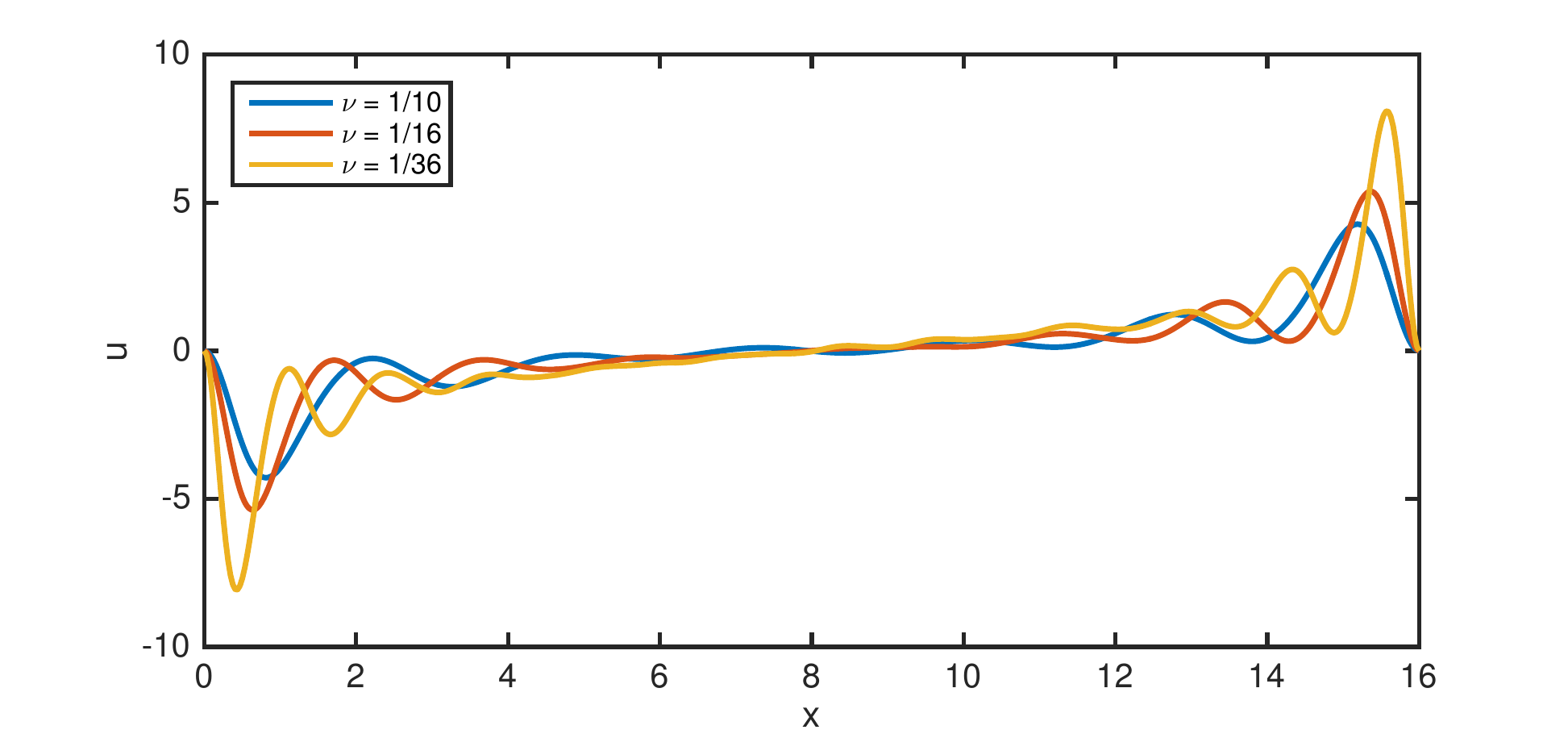}
          \end{subfigure}
          \begin{subfigure}[h]{.7\textwidth}
                \includegraphics[width=\linewidth]{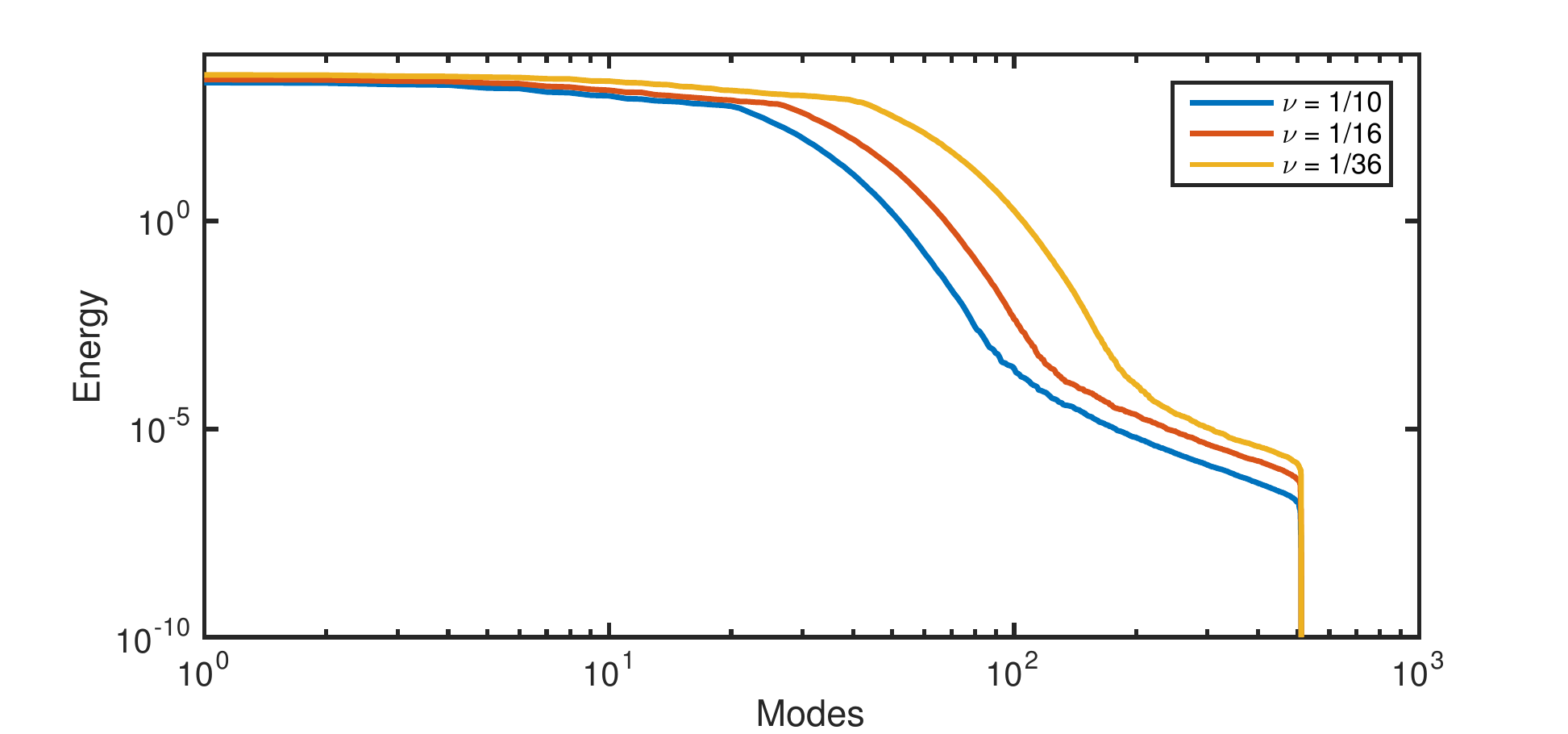}
          \end{subfigure}
          \caption{Comparison of (top) temporal average $\overline{u}=\frac{1}{T}\int_0^T
u \; dt$ and (bottom) energy spectrum $E_k$ vs. $k$ for different values
of $\nu$; systems with lower $\nu$ (more chaotic) have greater average oscillations
near the Dirichlet boundary conditions and slower drop-off in their energy
spectra.}
          \label{fig:KSspectrummean}
\end{figure}

\subsubsection{Dimension Reduction: Principal Component Analysis}

Applying our proposed method for constructing a data-based model, dimension
reduction is first performed on the 512-dimensional K-S system data. Here
we employ the most basic linear reduction method - principal component analysis
(PCA). In particular we reduce the system state as 
\begin{equation}
        \mathbf{a} = \mathbf{W}_d^T\tilde{\mathbf{u}}
\end{equation}where $\mathbf{W}_d$ is the $d$ eigenvectors of the covariance
matrix $\mathbf{C_{uu}}$ corresponding to the $d$ largest eigenvalues.
The tilde reflects the fact that the mean
state $\overline{\mathbf{u}}$ must be subtracted from data before the transformation,
i.e. $\tilde{\mathbf{u}} = \mathbf{u}-\overline{\mathbf{u}}$. Such a reduced
representation ensures that as much variance is retained in its $d$ components
as possible when limited to using a linear projection. The spectrum (eigenvalues of the covariance matrix) as well as the mean for different values of $\nu$ are shown in Figure \ref{fig:KSspectrummean}.

\subsubsection{Model Simulations and Results}

\begin{figure}[H]
        \centering
        \begin{subfigure}[h]{.425\textwidth}
                \includegraphics[width=\linewidth]{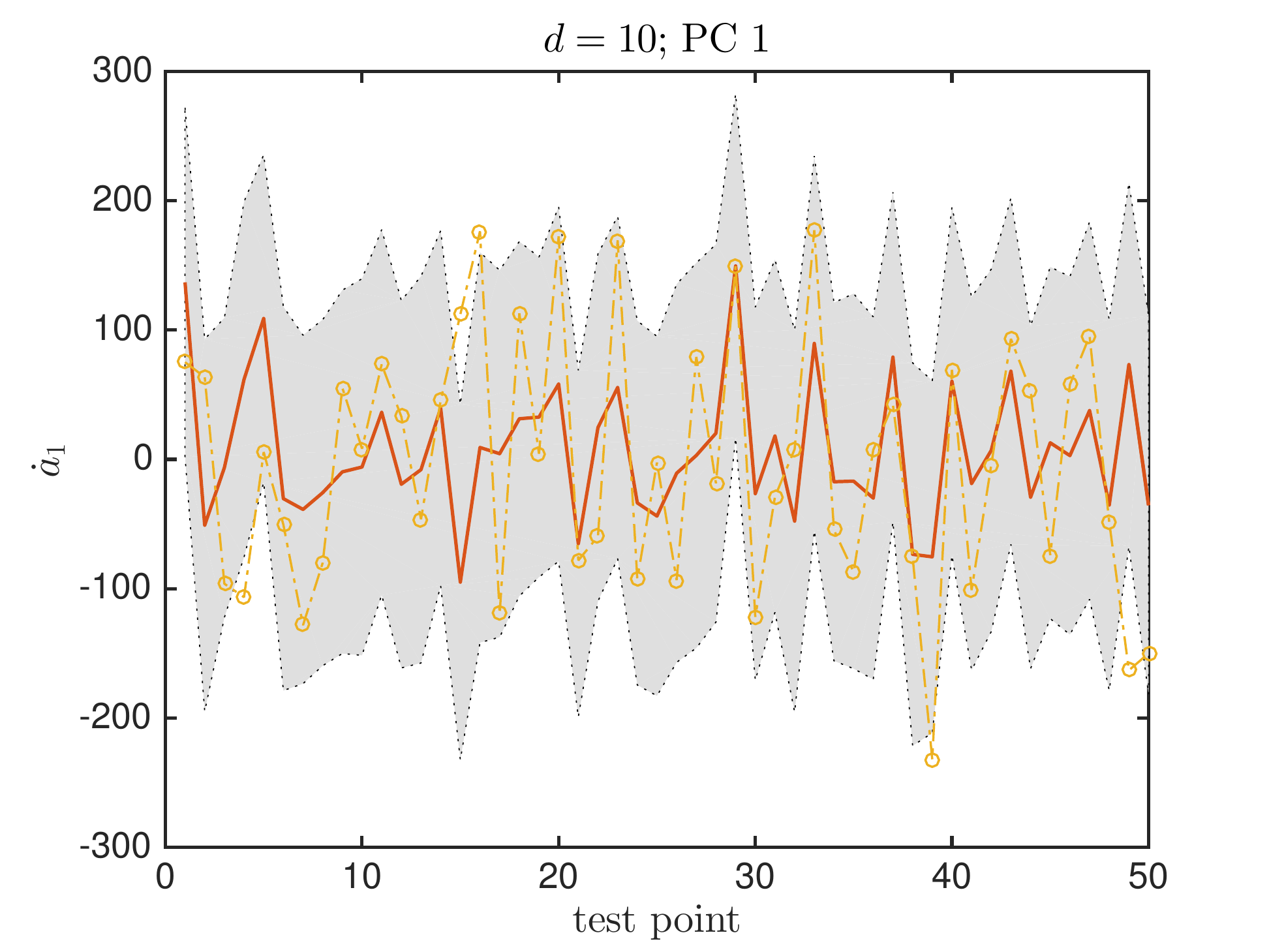}
          \end{subfigure}
          \begin{subfigure}[h]{.425\textwidth}
                \includegraphics[width=\linewidth]{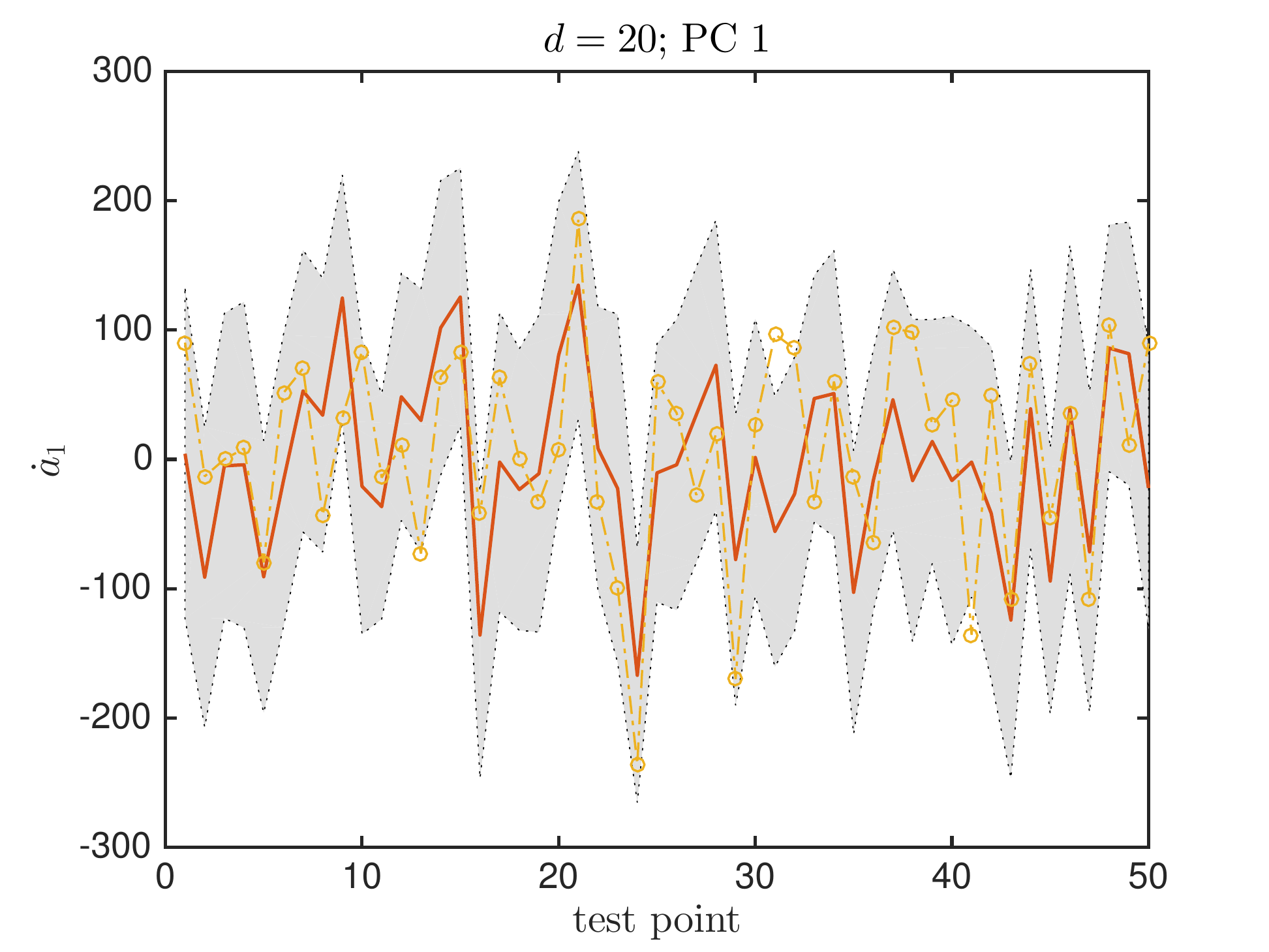}
          \end{subfigure}
          \caption{Validation plots: predicted dynamics of the first
principal component $\dot{a}_1$ with (left) $d=10$ and (right) $d=20$; predicted mean (red solid) is compared
with the true test point dynamics (yellow dashed); shaded regions capture
$2\sigma_1$ on both sides of $\dot{a}_1$. Horizontal axis corresponds to index of the test point.} 
          \label{fig:KSvalidation}
\end{figure}

We train $d$ GPR models by learning
the appropriate hyperparameters using methods described previously. Figure
\ref{fig:KSvalidation} shows some example GPR model predictions for the dynamics
of the first principal component in 10- and 20-dimensional reduced space
using the optimized hyperparameters respectively. When the model is constructed
on a 20-dimensional reduced-order space, much smaller uncertainties are predicted
for the dynamics because the unmodeled modes contain smaller percentages
of the overall system energy and thus less prominent effects.

The model parameters for the MSM and blended models are also calculated using the same data series. All models are solved by numerical integration with the
Euler-Maruyama method and a time step of $\Delta t = 0.02$. For each run we evolve 50 independent paths, whose initial conditions are drawn from a Gaussian distribution with variance
1 in all dimensions of the original space centered around the initial condition.
The integrations are carried out to $T = 5$.

\begin{figure}[p]
        \centering
        \begin{subfigure}{\textwidth}
                \includegraphics[width=\linewidth]{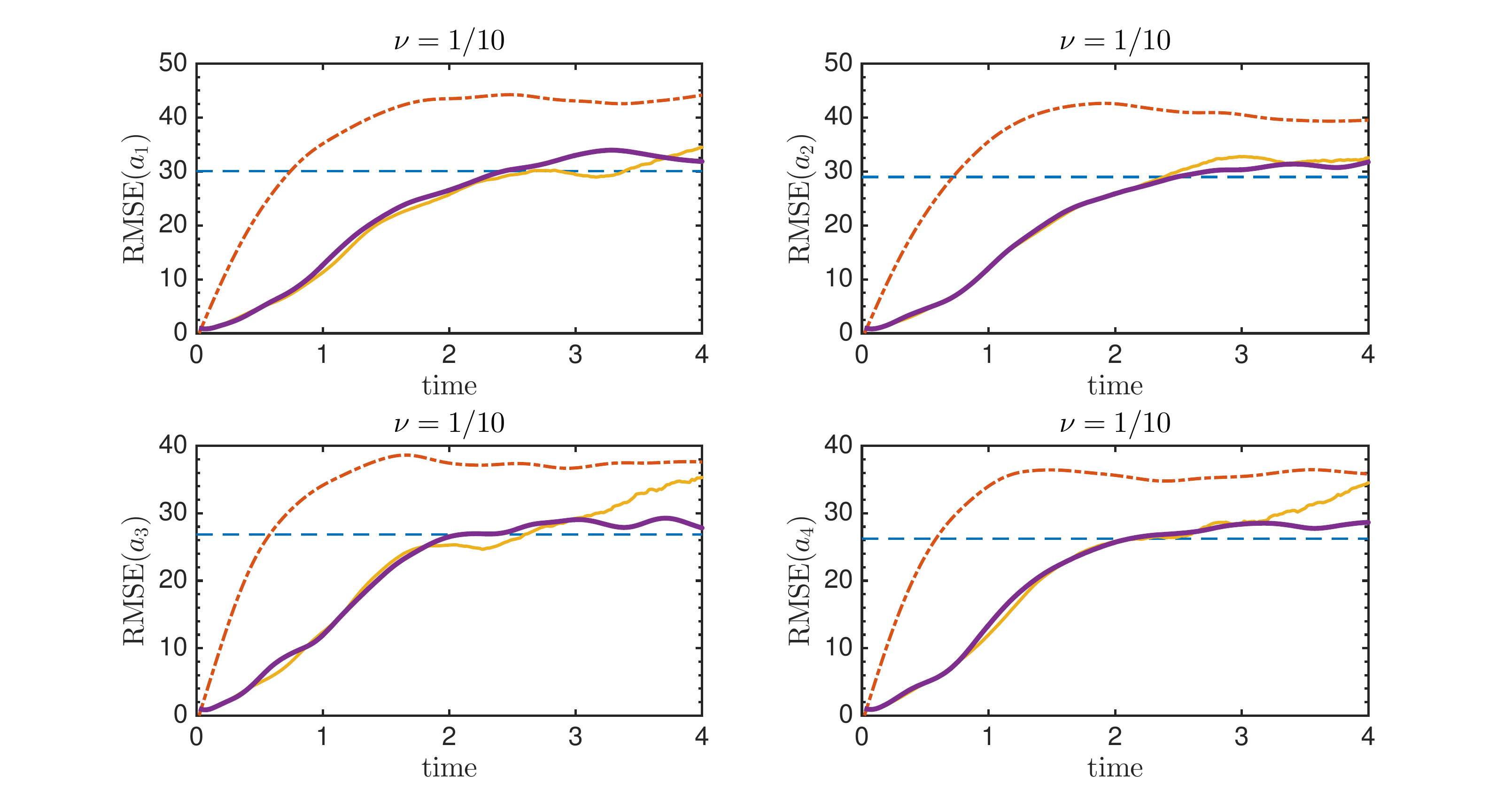}
                \captionsetup{justification=centering}
                \caption{}
        \end{subfigure}
        \begin{subfigure}{\textwidth}
                \includegraphics[width=\linewidth]{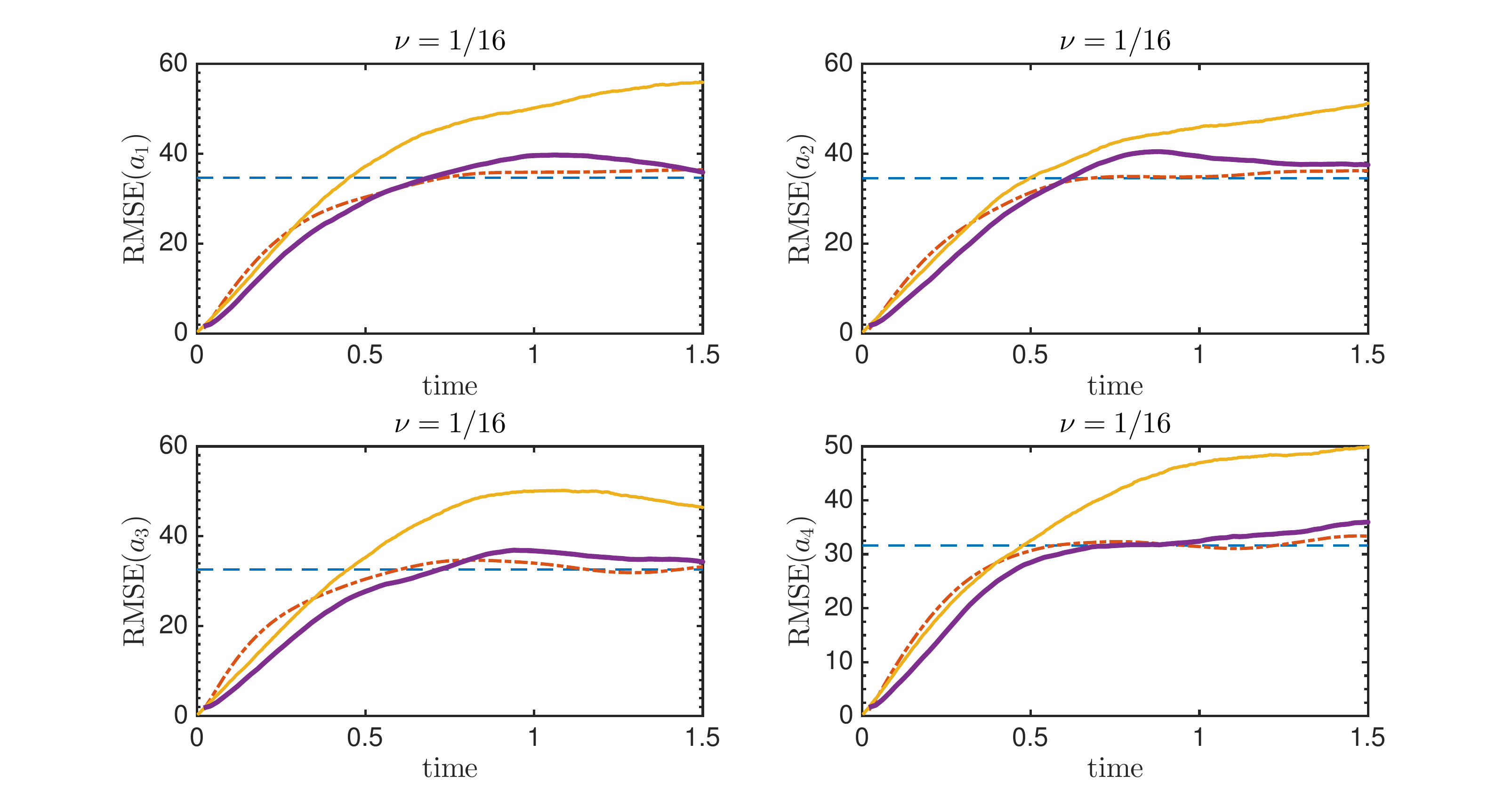}
                \captionsetup{justification=centering}
                \caption{}
        \end{subfigure}
        \caption{RMSE comparison for the K-S system with (a) $\nu = 1/10$ and (b) $\nu = 1/16$. Standard deviation of the attractor (blue dashed line); MSM (red dashed line); 20 principal components GPR (yellow solid line); blended GPR-MSM (purple solid thick line). All results are obtained by averaging over 1000 test initial conditions.}
        \label{fig:RMSE_KS}
\end{figure}

To evaluate the performance of the scheme we compare the model prediction with the true state. As truth we consider the trajectory evolved from the
initial condition $\mathbf{u}_0$ with the exact equations (\ref{eq:KSFD}). The true state in physical space
is then projected onto the most energetic principal components to give the
true reduced-order state, denoted by $\mathbf{a}^{\text{t}}$. The comparison is
performed for the same initial conditions. As a metric, the standard root
mean squared error (RMSE) is used

\begin{equation}
        \text{RMSE} = \sqrt{\frac{1}{V}\sum\limits_{i=1}^{V}\left({a^{\text{f}}_l}^{(i)}-{a^{\text{t}}_l}^{(i)}\right)^2},
        \label{eq:RMSE}
\end{equation}where subscript $l$ denotes the forecast/truth for the $l$th
leading principal component; $V$ is the total number of initial conditions
tested. The RMSE is calculated at successive time instances to generate error
curves characterizing the evolution of error with time. Figure \ref{fig:RMSE_KS}
shows the reduced-space RMSE curves for predicting the K-S system at $\nu
= 1/10$ and $\nu = 1/16$.  In each plot, we have included four error curves:
(1) invariant measure corresponding to the error for using the (constant) attractor
mean as the prediction, (2) MSM forecast, (3) reduced GPR model with 20 principal
components (4) mixed GPR-MSM with 20 principal components. $\nu = 1/10$ corresponds
to a less chaotic regime, where 20 principal components account for approximately
85\% of the total variance. For $\nu = 1/16$, the system is much more
chaotic and the attractor has a higher intrinsic dimension. In this case,
the first 20 principal components account for less than 80\% of the total
variance.

From the simulation results we observe that for the weakly chaotic regime $\nu = 1/10$ the reduced-order GPR performs better than the much less expensive MSM. However, the error does not stabilize and shows signs of diverging in the long run. This problem is not present for the blended scheme
where the error remains bounded as in MSM and converges properly to the invariant
measure; this is not obvious in the plots due to the extremely
slow damping coefficients $c_i$. In the more chaotic regime $\nu = 1/16$, trajectories are much more sensitive
to initial conditions. As a result, the same 20 principal components are
expected to capture less percentage of the overall dynamics. This is manifested
in the error curve as it grows much faster to the invariant measure and the
prediction performance of the reduced GPR model is now comparable to that of
the MSM. From these results it is clear that a low intrinsic dimensionality is important in order for the developed data-driven scheme to perform well. For highly turbulent systems where this is not the case the less expensive MSM models is the better option.

\begin{figure}[H]
        \centering
          \includegraphics[width=0.7\linewidth]{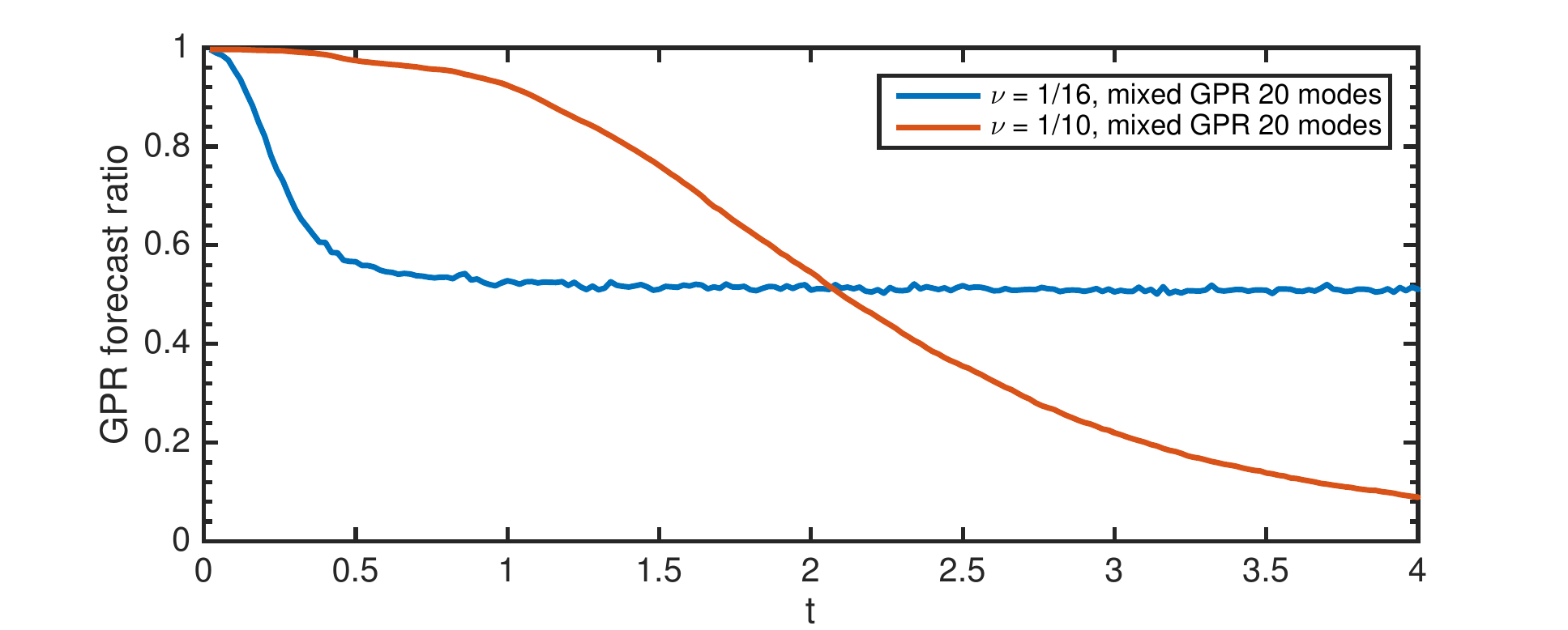}

          \caption{Percentage of ensemble members forecasted with GPR over time for
$\nu = 1/10$ and $\nu = 1/16$.}
          \label{fig:KS_RMSEnGPR}
\end{figure}
Figure \ref{fig:KS_RMSEnGPR} shows the average value of $\chi$ vs. time. The proportion of ensemble members
forecasted with GPR drops off much faster in the more chaotic regime due
to more trajectories approaching edges of the attractor more quickly. Meanwhile,
the steady state mixture ratio for $\nu = 1/10$ is smaller due to the slow
damping coefficients: it takes longer for MSM to drive a trajectory back
than for GPR to diffuse it away from the attractor. Hence, trajectories are
more likely to be governed by MSM in the long term. \\

\subsection{Lorenz 96 System}

\begin{figure}[H]
        \centering
          \begin{subfigure}[h]{.40\textwidth}
                \includegraphics[width=\linewidth]{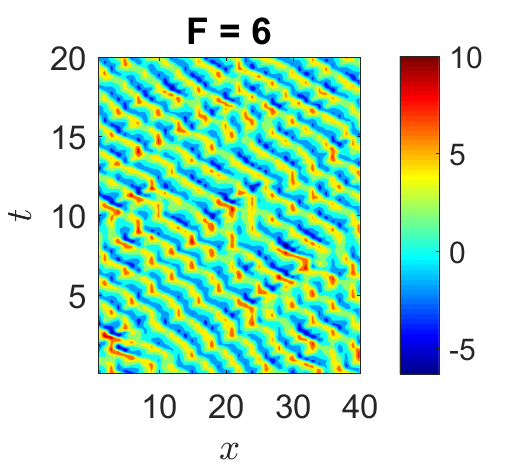}
          \end{subfigure}
          \begin{subfigure}[h]{.40\textwidth}
                \includegraphics[width=\linewidth]{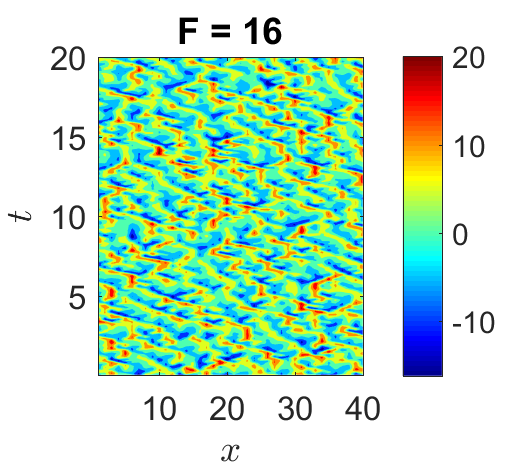}
          \end{subfigure}
          \caption{Contour plots for the L96 system exhibiting different levels of chaos as $F$ changes. $F=16$ system is much more turbulent than $F=8$ system}
          \label{fig:L96contours}
\end{figure}

In the second application we investigate the predictions skill of the developed scheme on the Lorenz 96 (L96) system, originally developed to crudely model the large scale behavior of the midlatitude atmosphere \cite{Lorenz96}. L96 is governed by the following system of nonlinear ordinary differential equations:

\begin{equation}
        \frac{dX_j}{dt} = (X_{j+1}-X_{j-2})X_{j-1} - X_{j} + F,
        \label{eq:L96}
\end{equation}
where $X_j$, $j = 0,...,J-1$, represent the "atmospheric variable" discretized spatially. Following \cite{Lorenz96}, we use $J = 40$. Periodic boundary conditions are applied. The system consists of a nonlinear advective-like term $(X_{j+1}-X_{j-2})X_{j-1}$, a linear dissipative term $- X_{j}$ and an external forcing term $F > 0$. The interactions amongst these terms conserve the total energy of the system and keep all $X_j$ always bounded. The system has a trivial equilibrium solution at $X_0 = ... = X_{J-1} = F$, which is unstable for sufficiently large values of the forcing parameter $F$. As we increase the value of $F$, the system moves from weakly chaotic regimes to fully turbulent \cite{sapsis_majda_mqg}, as shown in the contour plots in Figure \ref{fig:L96contours}. 

Data in this application is simulated by integrating (\ref{eq:L96}) from a single initial condition randomly perturbed from the equilibrium solution by a small amount for 10000 time units. Four sets of data are generated corresponding to $F = 4$, $F = 6$, $F = 8$ and $F = 16$. In all cases, a 4-step Runge-Kutta method is used with a time step of $\Delta t = 0.01$. 60\% of data is used for training and the rest is used for testing.

\subsubsection{Dimension Reduction: Fourier Analysis and Truncation}

Following \cite{majda_info} we pre-process the data by applying the following re-scaling and shifting, so that transformed variables $\tilde{X}_j$ have zero mean and unit energy (defined as $\frac{1}{2}\sum_j\tilde{X}_j^2$):

\begin{equation}
        X_j = \overline{X} + E_{p}^{1/2}\tilde{X}_j \;\; \mbox{and} \;\; t = E_p^{-1/2}\tilde{t},
        \label{eq:L96rescale}
\end{equation}where $\overline{X}$ is the mean state and $E_p$ is the average variance in energy fluctuation calculated as:

\begin{equation}
        E_p = \frac{1}{2T}\sum\limits_0^{J-1}\int_{T_o}^{T_o+T} (X_j - \overline{X})^2.
        \label{eq:Ep}
\end{equation}

Substituting (\ref{eq:L96rescale}) into (\ref{eq:L96}), we arrive at the following rescaled L96 model:
\begin{equation}
        \frac{d\tilde{X}_j}{d\tilde{t}} = E_p^{-1}(F-\overline{X})+E_p^{-1/2}((\tilde{X}_{j+1}-\tilde{X}_{j-2})\overline{X}-\tilde{X}_j)+(\tilde{X}_{j+1}-\tilde{X}_{j-2})\tilde{X}_{j-1}.
        \label{eq:nonDimL96}
\end{equation}

\begin{figure}[t]
        \centering
        \includegraphics[width=.8\linewidth]{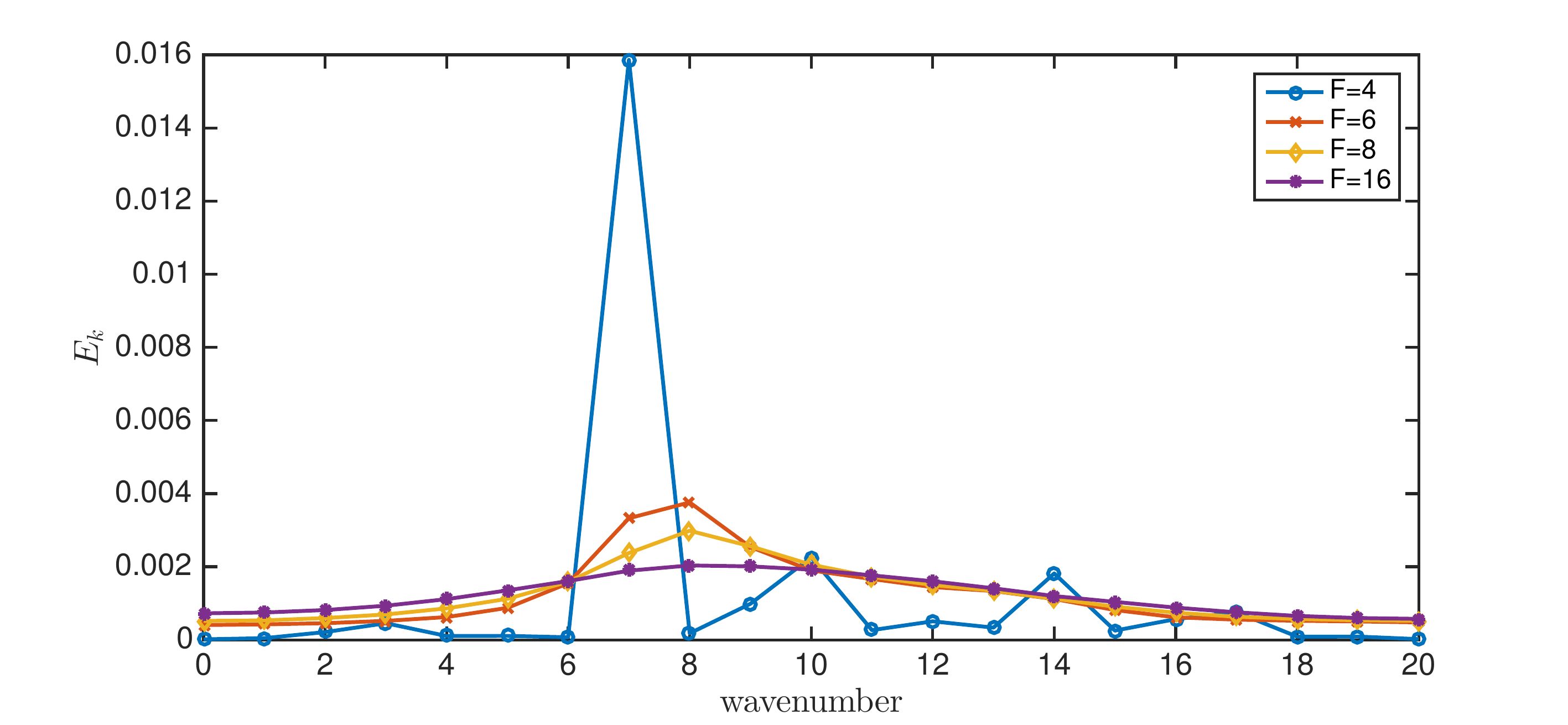}
        \caption{Energy Spectrum $E_k$ vs. $k$ for $F = 4$, $F = 6$, $F = 8$ and $F = 16$.}
        \label{fig:EspecAll}
\end{figure}

We then define the discrete Fourier transform and its inverse on the rescaled variables:

\begin{equation}
        \hat{X}_k = \frac{1}{J}\sum\limits_{j=0}^{J-1}\tilde{X}_je^{-2\pi ikj/J}, \;\;\; \tilde{X}_j = \sum\limits_{k=0}^{J-1}\hat{X}_ke^{2\pi ikj/J},
        \label{eq:dft}
\end{equation} while the energy of each Fourier mode $\hat{X}_k$ is defined as
\begin{equation}
        E_k \equiv Var(\hat{X}_k) = \overline{(\hat{X}_k(\tilde{t})-\overline{\hat{X}}_k)(\hat{X}_k(\tilde{t})-\overline{\hat{X}}_k)^*}. 
        \label{eq:var1}
\end{equation}
The resulting energy spectra $E_k$ for Fourier modes $k = 0,...,20$ under different forcing is shown in Figure \ref{fig:EspecAll}. We observe that L96 has very different energy levels in its Fourier modes, especially in the weakly chaotic regimes. Therefore, a simple and natural way to construct a reduced-order model is to use a few of its most energetic Fourier modes to form a truncated low-dimensional representation of the whole system and model the effects of the ignored modes stochastically. The most energetic modes for each forcing is summarized in Table \ref{tab:modes}. Note that $F = 6$, $F = 8$ and $F = 16$ energy spectra share the same 6 most energetic Fourier modes in slightly different orders. Note that the most energetic modes may not necessarily be the most dynamically relevant ones \cite{Crommelin2004}. However, here we focus on quantifying the approximation properties of the proposed scheme within a given subspace; the selection of this subspace is a different issue, which is problem- and context-dependent.  

\begin{table}[t]
\centering
\begin{tabular}{c|c}
        \hline
        Forcing Regime & Selected Wavenumbers $k$ (ordered by energy $E_k$) \\ \hline
        F = 4           & 7, 10, 14, 9, 17, 16                \\
        F = 6           & 8, 7, 9, 10, 11, 6                 \\
        F = 8           & 8, 9, 7, 10, 11, 6                 \\
        F = 16          & 8, 9, 10, 7, 11, 6                 \\ \hline
\end{tabular}
\caption{Most energetic Fourier modes used to construct reduced-space GPR models in each forcing regime.}
\label{tab:modes}
\end{table}

\subsubsection{Model Simulations and Results}

The six Fourier modes with the highest energy, $E_k$, in each regime are used as the reduced-order representation of the system state. Since the modes are complex-valued, the reduced space is in fact 12-dimensional. Independent GPR models are constructed for the dynamics of both the real and imaginary parts of the top six Fourier modes. 

\begin{figure}[h]
        \centering
        \includegraphics[width=\linewidth]{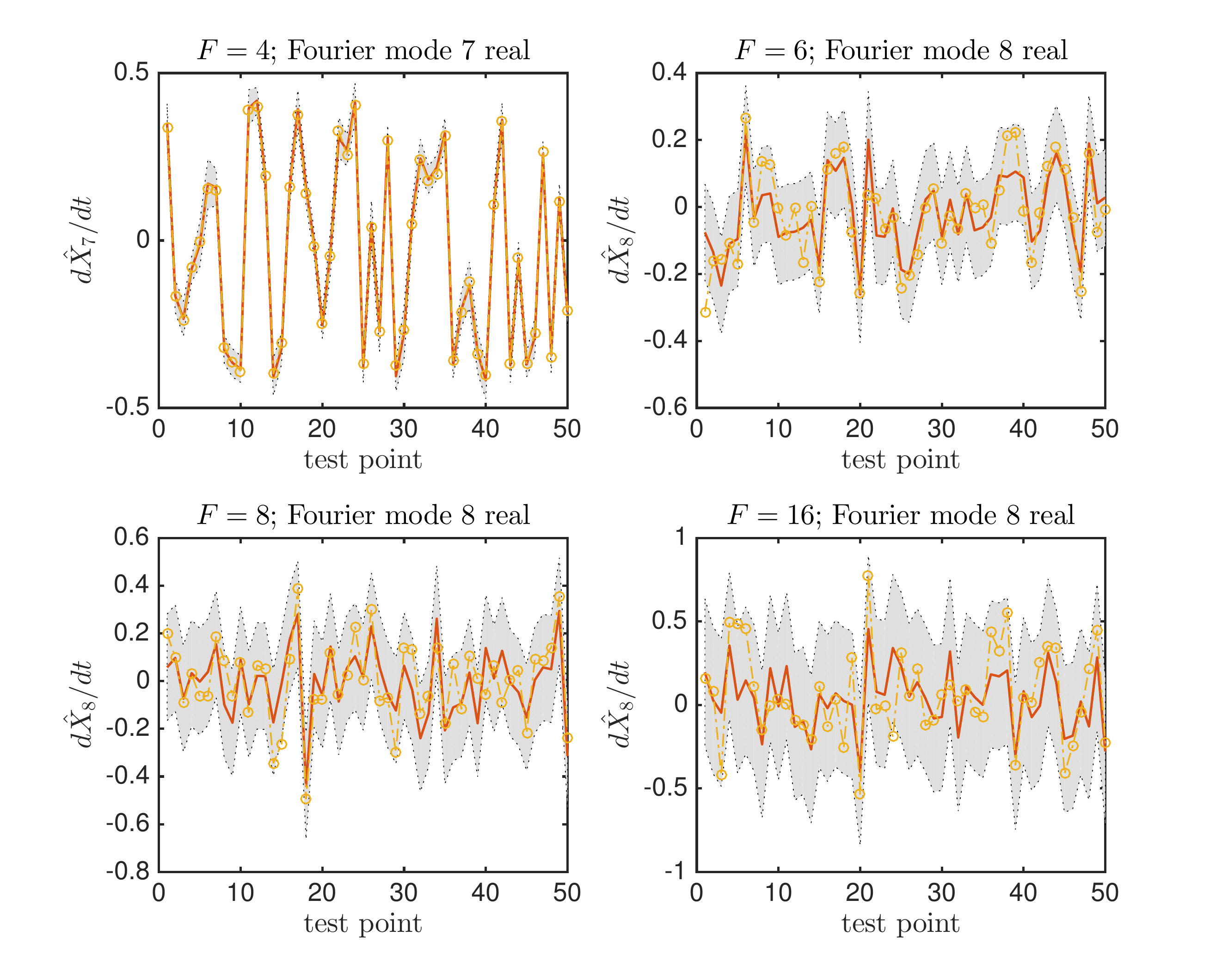}
        \caption{Validation plots for the dynamics of the most energetic mode of $F = 4$, $F = 6$, $F = 8$ and $F = 16$ with the GPR model (red solid), true test point dynamics (yellow dashed) and region within two standard deviations from the predicted mean (shaded). Predictions for smaller $F$ is more accurate with smaller error variances.}
        \label{fig:GPRvalidation}
\end{figure}

Figure \ref{fig:GPRvalidation} shows some examples of predicted dynamics for the real part of the most energetic Fourier mode in each forcing regime, along with the true dynamics estimated from finite differences. For $F = 4$, the dynamics can be predicted almost perfectly with very little error. As $F$ increases, the predicted dynamics remain close to the true value and well encapsulated within the two standard deviation interval. However, the predictions come with increasingly larger uncertainties as the unmodeled modes contain more energy and have bigger impact on the dynamics of the modeled modes. This is a similar pattern also observed in the K-S system.

\begin{figure}[p]
        \centering
        \begin{subfigure}{\textwidth}
                \includegraphics[width=\linewidth]{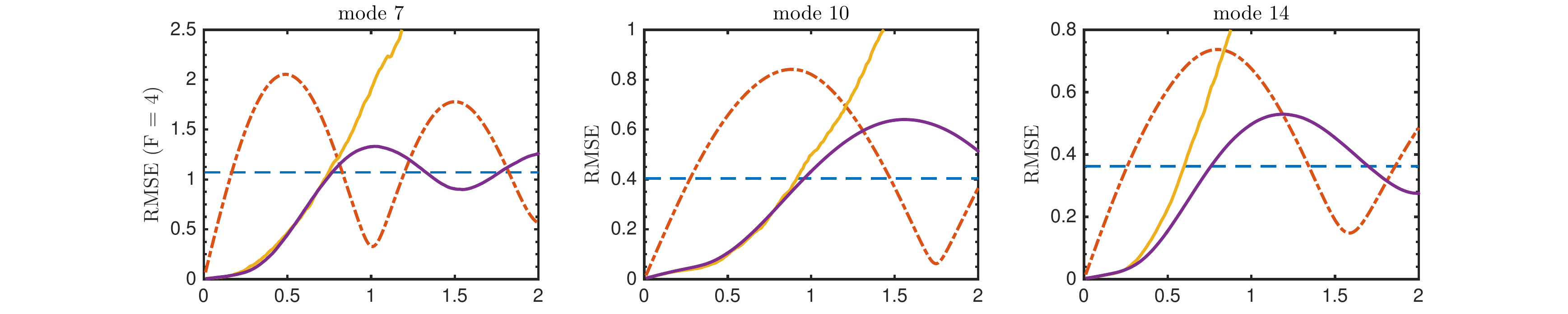}
        \end{subfigure}
        \begin{subfigure}{\textwidth}
                \includegraphics[width=\linewidth]{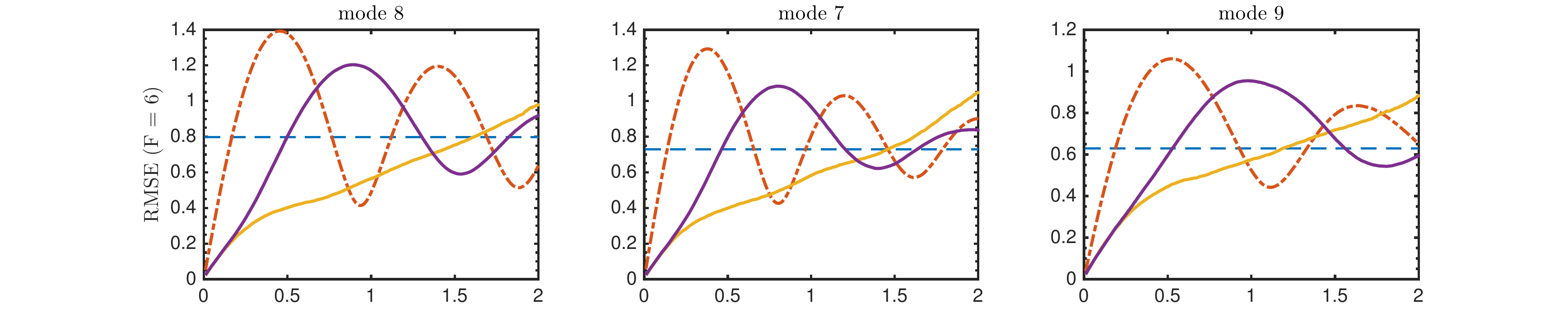}
        \end{subfigure}
        \begin{subfigure}{\textwidth}
                \includegraphics[width=\linewidth]{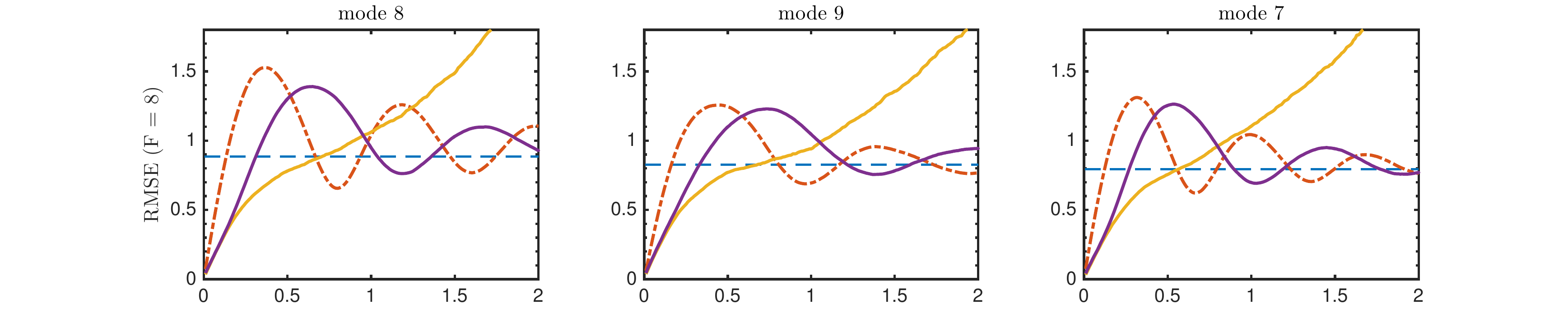}
        \end{subfigure}
        \begin{subfigure}{\textwidth}
                \includegraphics[width=\linewidth]{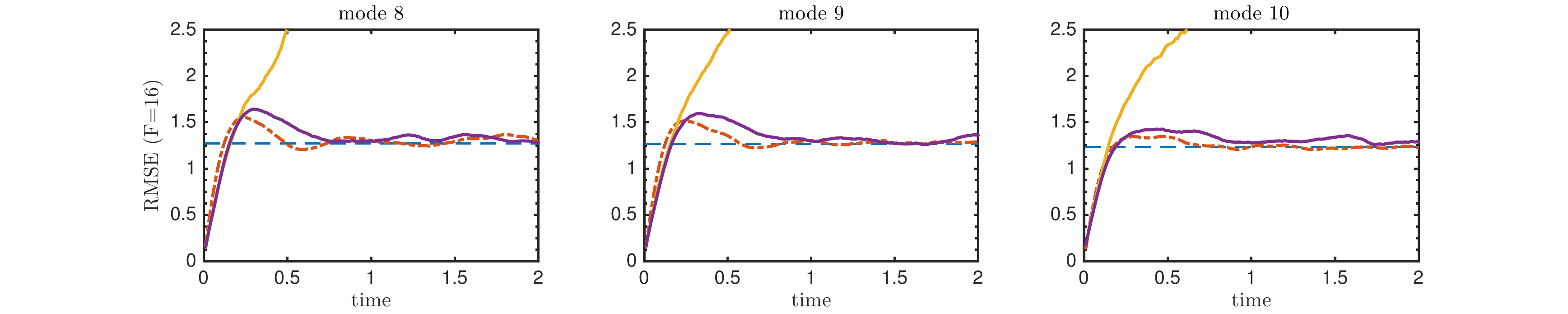}
        \end{subfigure}

        \caption{RMSE comparison of the three most energetic modes for L96 in different forcing regimes (units for both horizontal and vertical axes correspond to those before rescaling (\ref{eq:L96rescale})). Standard deviation from the attractor mean (blue dashed line); MSM (red dashed line); GPR (yellow solid line); blended GPR-MSM (purple solid thick line). All results are obtained by averaging over 1000 test initial conditions.}
        \label{fig:RMSE_L96}
\end{figure}

Using these GPR models, reduced space dynamics are forecasted for 1000 initial conditions randomly drawn from the test data set. Each initial condition is made Gaussian with covariance $0.1I$ in the original space. The true state is obtained by running an ensemble using the model equations (\ref{eq:L96}) and then applying rescaling and Fourier transform. The RMSE between the predicted mean state and the true mean state is calculated using definition (\ref{eq:RMSE}), with the exception that square of the complex absolute value is used instead. Results for MSM and blended models are also presented for comparison. The results are shown in Figure \ref{fig:RMSE_L96}. 

%

For $F = 4$, the RMSE of the reduced-order GPR prediction is much lower compared to that of MSM method in the short term. However, due to chaos in the system, error is amplified quickly and grows beyond the variance of the system. For $F = 6$, the reduced-order GPR still produces significantly better short-term prediction than the MSM, but the gap is visibly smaller than in the less chaotic $F = 4$ case. As forcing is increased, the gap further closes and the short-term performance of the reduced GPR and MSM becomes comparable to each other. This is a reasonable trend taking into account that the six modeled modes contain a smaller proportion of the total energy as the systems becomes more chaotic. The dynamical effect of the unmodeled modes become increasingly significant, to the extent that makes them collectively more important than the modeled modes. Thus, the diffusion term in (\ref{eq:model3}) dominates and the model behaves similarly to MSM. Hence, taking into account the low computational complexity, MSM is the appropriate model to use in very turbulent regimes. To increase the power of the reduced GPR approach in these highly turbulent regimes, a larger number of Fourier modes is required.

\begin{figure}[t]
        \centering
        \includegraphics[width=.75\linewidth]{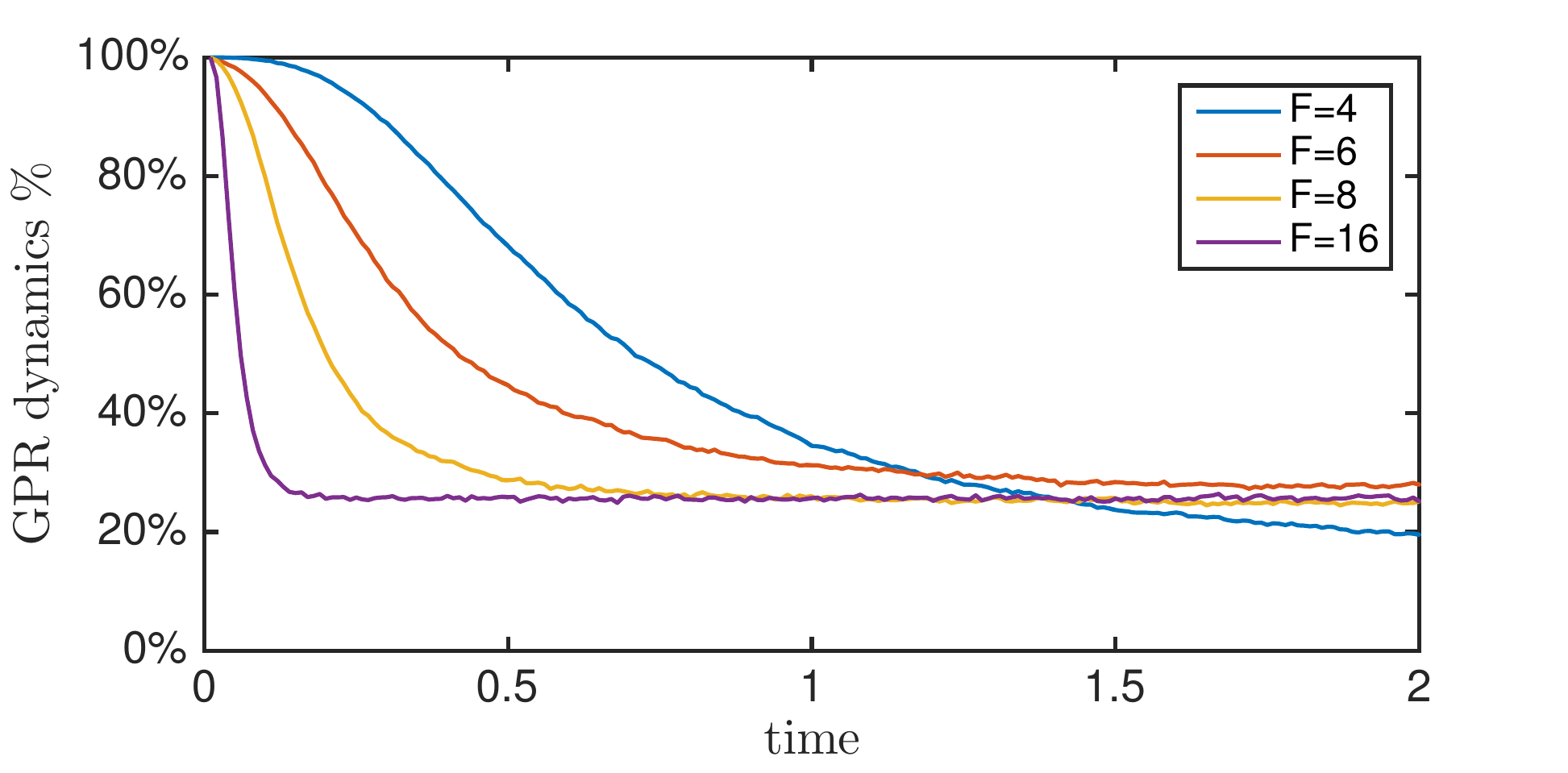}
        \caption{Percentage of ensemble members using GPR dynamics over time for different forcing regimes in L96. Average is performed over 500 initial conditions.}
        \label{fig:mixedProportions}
\end{figure}

The results also showcase the mixture model as an effective middle ground between the reduced GPR and MSM forecast, possessing both the short-term accuracy of the former and the long-term stability of the latter. Figure \ref{fig:mixedProportions} shows the percentage of the ensemble for which GPR dynamics is used as time evolves. We observe that the rate of drop in this percentage increases with $F$ and is consistent with the rate of increase in the average RMSE for the reduced GPR forecast.

\subsection{A barotropic climate model}

\begin{figure}[t]
        \centering
        \includegraphics[width=0.88\linewidth]{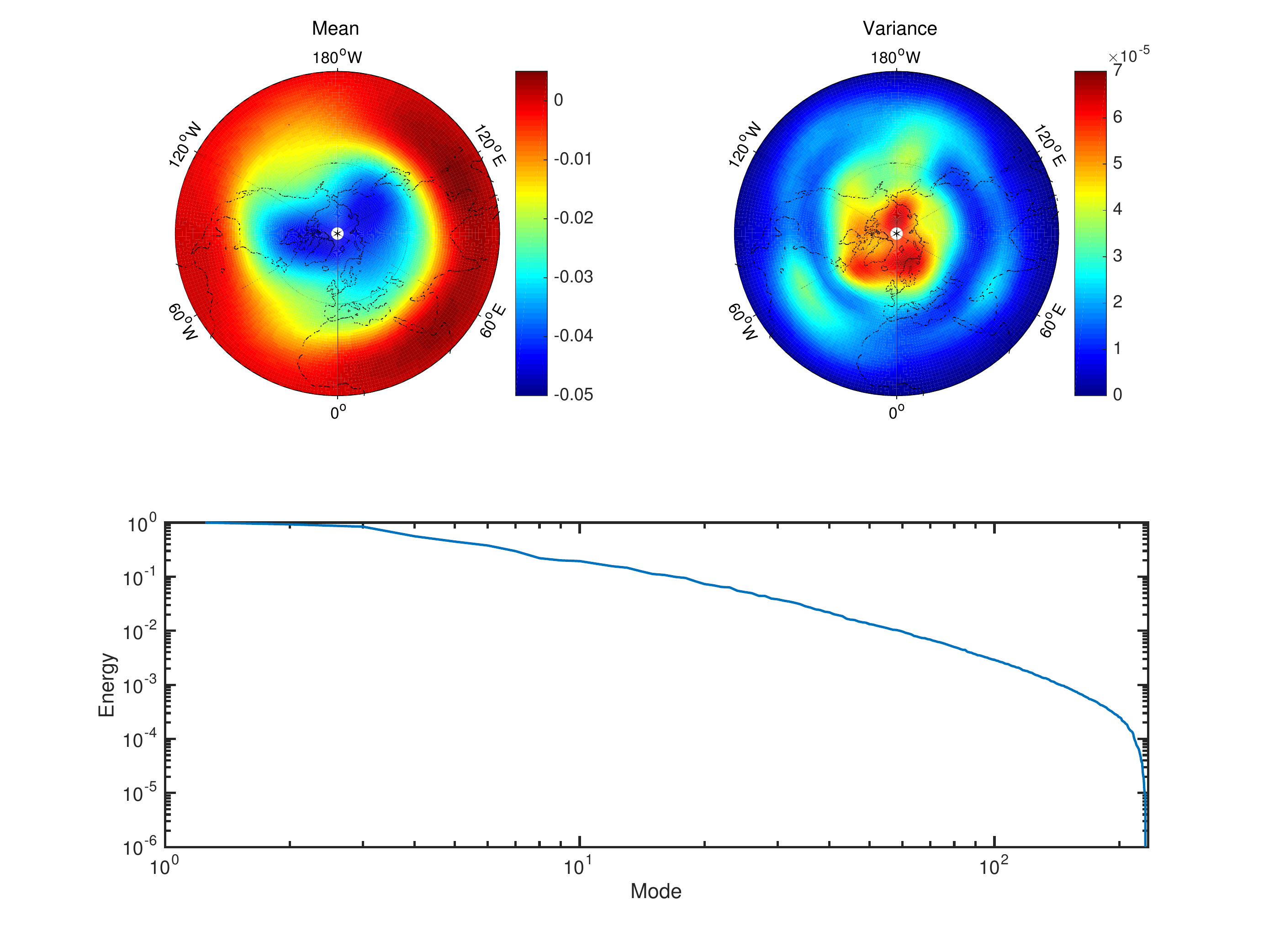}
        \caption{The mean (top left), variance (top right) and energy spectrum (bottom) of the T21 barotropic model in statistical steady state.}
        \label{fig:T21_meanvarspectrum}
\end{figure}

We now investigate the effectiveness of the proposed method on a spectral barotropic model on a spherical earth with realistic orography \cite{fran_majd}. This model has its forcing parameters calculated from observations so that the climatological mean and low-frequency variability is realistic. The model equation is given by

\begin{equation}
        \frac{\partial \zeta}{\partial t} = -J(\psi,\zeta+f+h)-\frac{\zeta}{\tau}+K\Delta^3\zeta+F
        \label{eq:T21model}
\end{equation}
where $\zeta$ represents relative vorticity, $f=2\Omega\sin\theta$ is the Coriolis parameter ($\Omega$ is the angular velocity of the earth), $h$ is the Ekman damping coefficient, $K$ is the coefficient of the scale-selective damping, and $F$ is the external time-dependent forcing. Furthermore, $\psi$ is the corresponding stream function such that $\zeta = \Delta\psi$. Under appropriate nondimensionalization, this equation has unit length equal to the radius of the earth and unit time equal to the inverse of the angular velocity of the earth. The Jacobi operator $J(a,b)$ is defined as

\begin{equation}
        J(a,b) = \left(\frac{\partial a}{\partial \lambda}\frac{\partial b}{\partial \mu}-\frac{\partial a}{\partial \mu}\frac{\partial b}{\partial \lambda}\right)
        \label{eq:T21J}
\end{equation}
where $\mu$ is the sine of the latitude and $\lambda$ is the sine of the longitude. The nondimensionalized orography $h$ is related to the real orography $h'$ by $h=A_0h'/H$ where $A_0 = 0.2$ defines the strength of the wind blowing on the surface of the orography and $H = 10$km is a height scale \cite{selten95}. The model is truncated at T21. By restricting the spectral model to modes whose zonal wavenumber and total wavenumber sum up to even numbers, a model of hemispheric flow is obtained with a total number of 231 variables. The data used to set up the GPR and MSM models is acquired by integrating equation (\ref{eq:T21model}) for $10^5$ days after an initial spin-up period of 1000 days, using a fourth-order Adams-Bashforth integration scheme with a 45-min time step. The Ekman damping time scale is set to 15 days and the strength of the scale is selected such that wavenumber 21 is damped at a time scale of 3 days. The spatial domain (spherical surface) is discretized into a $D = 64\times32$ grid with equally spaced latitude and longitude. 80\% of data is randomly selected and used for training while the rest is used for testing. Figure \ref{fig:T21_meanvarspectrum} top shows the mean and variance of the statistical steady state.

\subsubsection{Dimension Reduction: Classical Multidimensional Scaling} 

For the dimension reduction portion of this problem, we use a generalized version of the classical multidimensional scaling (MDS) procedure. It is motivated by the idea of preserving scalar products, i.e. the lower-dimensional embedding for a data set should be created such that the original pairwise scalar products are preserved as much as possible. Hence, assuming the products are clearly defined in both the original and the reduced space\footnote{PCA is equivalent to MDS if the distance measures in both spaces are defined to be the scalar vector (dot) product; hence PCA is also called classical MDS}, MDS seeks to solve the minimization problem

\begin{equation}
        \min\limits_{\mathbf{y}_1,...,\mathbf{y}_N} \sum\limits_{i<j}(s_{\boldsymbol{\zeta}}(i,j)-s_{\mathbf{y}}(i,j))^2
        \label{eq:MDSobj}
\end{equation}
where $s_{\boldsymbol{\zeta}}$ and $s_{\mathbf{y}}$ respectively denotes the product function defined in the original $\boldsymbol{\zeta}$ space and the reduced $\mathbf{y}$ space. $(i,j)$ are indices of the data between which products are calculated. This objective function minimizes the total squared error between pairwise products. When $s_{\mathbf{y}}$ is chosen to be the scalar vector (dot) product, the analytical solution to this optimization problem can be obtained. Let $[W]_{ij} = s_\mathbf{\boldsymbol{\zeta}}(i,j)$ be the Gram matrix, and its eigenvectors be sorted in descending order by absolute value: $|\kappa_1| \geq |\kappa_2| \geq ... \geq |\kappa_N|$. The optimal $d$-dimensional embedding for a training point $\boldsymbol{\zeta}_n$ under (\ref{eq:MDSobj}) can be written in terms of the eigenvectors of the $W$, $W\mathbf{v}_l = \kappa_l\mathbf{v}_l$, with $\mathbf{v}_l = [v_{1,l},...,v_{N,l}]^T$, as follows

\begin{equation}
        \mathbf{y}_n = 
        \begin{pmatrix}
                \kappa_1^{1/2}v_{n,1}\\
                \kappa_2^{1/2}v_{n,2}\\
                \vdots \\
                \kappa_d^{1/2}v_{n,d}
        \end{pmatrix}.
        \label{eq:MDSembed}
\end{equation}

\begin{figure}[t]
        \centering
        \includegraphics[width=.9\linewidth]{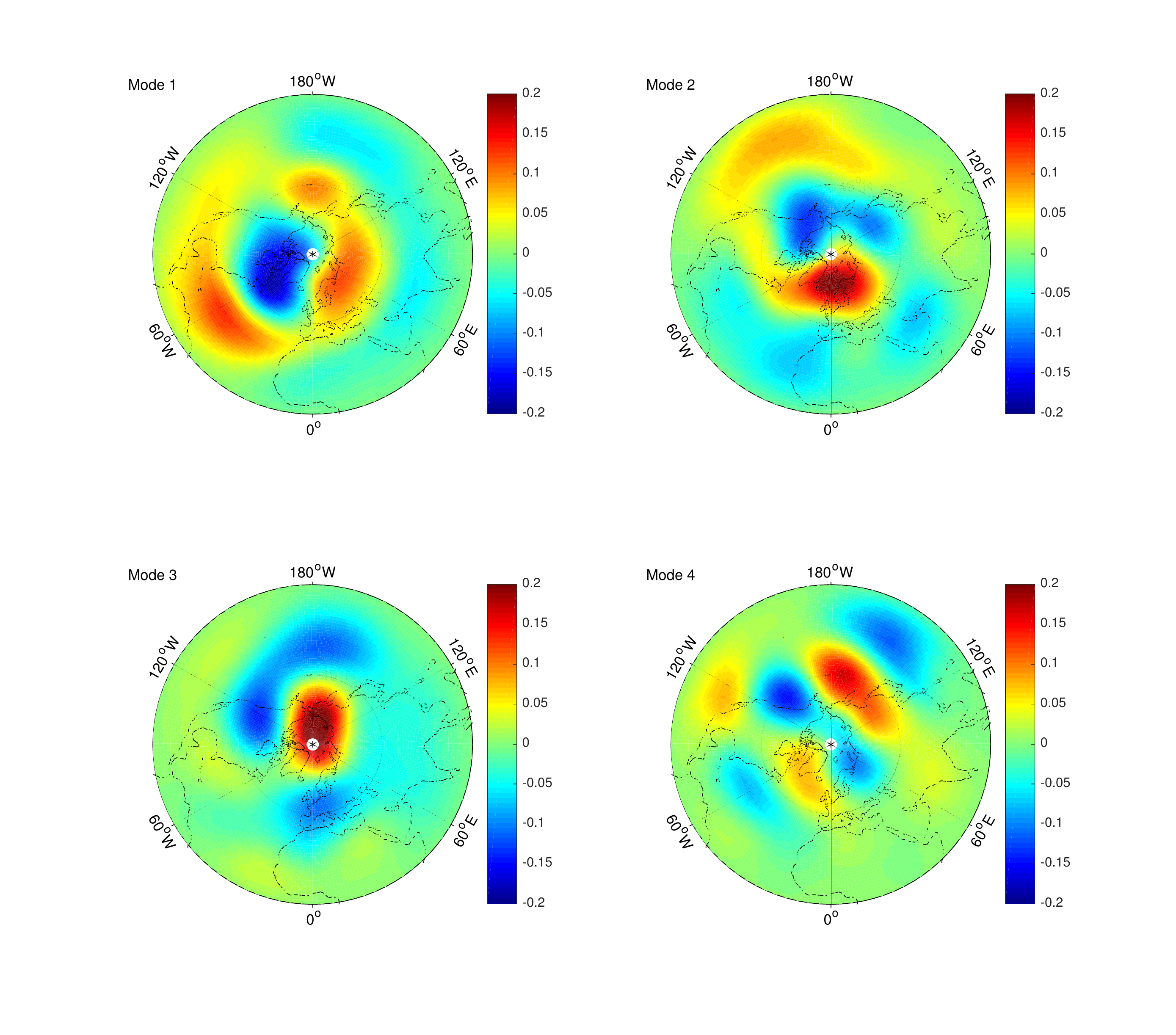}
        \caption{The four most energetic empirical orthogonal functions of the climatology.}
        \label{fig:T21_EOF4}
\end{figure}
Since equation (\ref{eq:MDSobj}) in matrix form can be also interpreted as finding the best low-rank approximation to the Gram matrix in terms of the Frobenius norm, the optimality of (\ref{eq:MDSembed}) can be proven by the Eckart-Young-Mirsky theorem. Specifically for this problem, we use the standard kinetic energy product as measures of proximity between states, which preserves the nonlinear symmetries of the dynamics for the system (\ref{eq:T21model}):

\begin{equation}
        s_\mathbf{\boldsymbol{\zeta}}(i,j) = \langle\zeta_i \cdot \zeta_j\rangle = \int_\mathcal{S}\nabla\psi_i\cdot\nabla\psi_j \;d\mathcal{S} = -\int_\mathcal{S}\zeta_i\psi_j \;d\mathcal{S} = -\int_\mathcal{S}\zeta_j\psi_i \;d\mathcal{S}.
        \label{eq:T21IP}
\end{equation}
The energy spectrum associated with this definition of $s_{\boldsymbol{\zeta}}$ is shown in the bottom Figure \ref{fig:T21_meanvarspectrum}. Note, however, that (\ref{eq:MDSembed}) only gives the optimal embedding for the $N$ training points used to construct the Gram matrix $W$. To calculate the embedding for a new point, it is convenient to first find the empirical orthogonal functions (EOFs) corresponding to each dimension of the reduced-order space

\begin{equation}
        \label{eq:EOF}
        \boldsymbol{\phi}_m = \sum\limits_{n=1}^{N} \kappa_{m}^{-1/2}v_{n,m}\boldsymbol{\zeta}_n, 
\end{equation}
where $m$ runs from 1 to $d$. The EOFs share the same dimension with $\boldsymbol{\zeta}$ and naturally ranked according to their energy level. In addition, they are orthogonal with respect to the energy product (\ref{eq:T21IP}), i.e. $\langle\boldsymbol{\phi}_{m_1},\boldsymbol{\phi}_{m_2}\rangle = \delta(m_1,m_2)$, where $\delta$ denotes the Kronecker-delta function. The first four EOFs are shown in Figure \ref{fig:T21_EOF4}. They account for 13.5\%, 11.4\%, 10.4\% and 7.1\% of the total energy respectively. Moreover, these EOFs bear resemblance to realistic climate patterns. For example, the first EOF is characterized by a center of action over the Arctic that is surrounded by a zonal symmetric structure in midlatitudes, similar to the Arctic Oscillation/Northern Hemisphere Annular Mode (AO/NAM) \cite{thom98}. The second, third and four EOFs are comparable to the East Atlantic/West Russia \cite{Barn87}, the Pacific/North America (PNA) \cite{wall81} and the Tropical /Northern Hemisphere (TNH) \cite{mo86} patterns respectively. Therefore, predictions for these EOFs have high practical significance because they are analogous to predicting corresponding climate patterns.

The calculated EOFs act as mode shapes and the components of $\mathbf{y}$ as the coefficients for these mode shapes, since it can be easily verified that $\boldsymbol{\zeta} = \sum_{m=1}^{N}y_m\boldsymbol{\phi}_m$ for a full expansion $d = N$.  Thus, using the orthogonality property, the reduced representation $\mathbf{y}^*$ for a new state $\boldsymbol{\zeta}^*$ can be obtained by taking inner products (\ref{eq:T21IP}) with the EOFs, i.e.

\begin{equation}
        \mathbf{y}^* = 
        \begin{pmatrix}
                \langle\boldsymbol{\zeta}^*,\boldsymbol{\phi}_1\rangle \\
                \langle\boldsymbol{\zeta}^*,\boldsymbol{\phi}_2\rangle \\
                \vdots \\
                \langle\boldsymbol{\zeta}^*,\boldsymbol{\phi}_d\rangle
        \end{pmatrix}
\end{equation}

\subsubsection{Model Simulations and Results}

\begin{figure}[t]
        \centering
        \includegraphics[width=\linewidth]{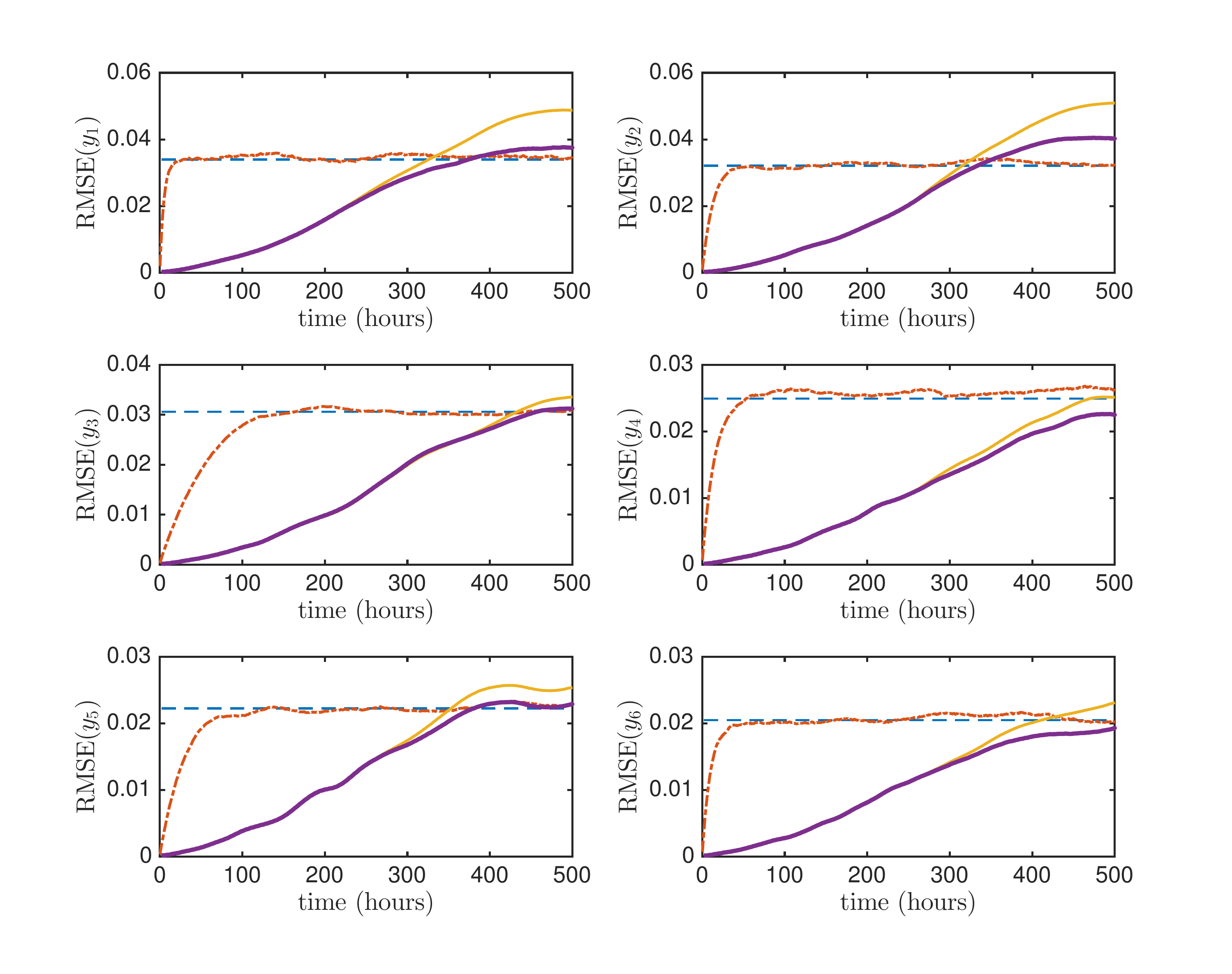}
        \caption{RMSE comparison for 6 most energetic modes in T21 system: standard deviation of the attractor (blue dashed line); MSM (red dashed line); GPR (yellow solid line); blended forecast (purple solid thick line); All results are obtained by averaging over 1000 test initial conditions.}
        \label{fig:T21_RMSE}
\end{figure}

Here we use $d = 30$, as $\boldsymbol{\phi}_d$ contains only about 3\% of $\boldsymbol{\phi}_1$'s energy. Similarly to the previous applications, 30 GPR models are trained for real-valued $\dot{y}_1,...,\dot{y}_{30}$. 1000 points are randomly picked from the attractor and used as the testing initial conditions. Centered around each initial condition, a Gaussian ensemble with a small variance ($1 \times 10^{-3}$) along each dimension is formed and marched forward using the reduced-order GPR, MSM and blended modeling approaches respectively. We then calculate the average RMSE of the predictions measured against the true states calculated using the true dynamics. The resulting error comparison is shown in Figure \ref{fig:T21_RMSE}. 

We observe that the reduced-order GPR approach in the short term significantly outperforms the MSM predictions: the GPR error curves generally take 300 to 400 hours to reach the standard error of the real attractor - at least three times longer than the 100 hours offered by the MSM forecast. This is because the reduced GPR approach takes better advantage of the inherent low-dimensional structure of the underlying attractor. It models the energy change for each mode due to nonlinear effects much more precisely. The considered climate model is much less turbulent than the Lorenz 96 system with $F \geq 16$, for which all models have comparable performance due to the strongly turbulent character of the attractor. In particular, the rates of change in the modeled modes are much higher at some locations than others. GPR performs better at differentiating these fast-growing regions of the attractor from the relatively steady ones while MSM only admits similar rates of change everywhere and for all times. Hence, when averaged over a larger number of initial conditions, the GPR-based forecasts have much better performance overall.

The introduction of mixture model in this case does not compromise the prediction performance of the GPR in the short term like in the $F = 6$ regime of L96 system. It helps effectively control the variance to match that of the true system in the statistical steady state.

\section{Conclusions}

We have formulated a reduced-order data-driven prediction method for chaotic dynamical systems using the Gaussian Process Regression (GPR) technique. The developed approach characterizes reduced-order dynamics in terms of a deterministic and a stochastic component. The deterministic component mainly embodies the dynamics due to the explicitly modeled dimensions/modes while the standard error represents dynamics uncertainty due to sources such as the unmodeled modes and interpolation errors. Based on these two components, a purely data-driven stochastic model is formulated in the reduced-order space and solved to provide a probabilistic forecast for the most important modes of the system. This modeling technique is highly generic and can adapt to a wide variety of nonlinear chaotic systems. 

For dynamical systems exhibiting different levels of turbulence/chaotic behaviors, comparison is carried out on the basis of the root mean squared error (RMSE) between trajectories forecasted with model dynamics and real dynamics. In addition, the mean stochastic model (MSM) method is also implemented as a benchmark method. Numerical experiments demonstrate that the GPR-based forecast has much lower errors provided that the attractor has low intrinsic dimensionality, i.e. the majority of the energy in the system is captured in the reduced dimensions whose dynamics are explicitly modeled. However, this condition becomes more difficult to satisfy as the system becomes more turbulent since energy tends to spread out in more modes than that can be efficiently included in the GPR models. As a result, the GPR dynamics are dominated by their stochastic components and prediction performance is comparable to that of the MSM. Moreover, long-term error of GPR forecast does not naturally converge to the standard error of the attractor. To resolve this issue we develop a blended GPR-MSM model so that GPR is only performed at locations enclosed by sufficient training data. The blended approach guarantees stable and consistent steady-state behavior matching that of the attractor.

There is a number of promising directions towards which studies can be extended. One possible such direction is to use adaptive modeling such that different hyperparameters and/or training data sets are used to make forecasts at different locations on the attractor. In this way, we can improve the prediction skill in rapidly-changing regions by using finer length scales and more data evidence while maintaining high accuracy with smaller training data sets and bigger length scales in smooth regions. Even though this comes at the expense of more computation costs, sparse GPR algorithms (see \cite{rasmusparse05}) can be used. Another direction is to take into account correlations in the prediction of dynamics between different variables from the reduced-order space, instead of modeling them independently of one another. Possible mechanisms for achieving this goal include using a correlated noise process and introducing correlations in the GP prior (see cokriging \cite{cressie93, perdikarisPRSA2015}). Although more involved theoretical development and supporting numerical simulations are needed, both directions have the potential to further upgrade the prediction skills of GPR dynamical models. Finally, the developed approach can form the basis for the combination of data-driven and adaptive order-reduction methods (e.g. combined with dynamically orthogonal equations, \cite{sapsis11a, sapsis_majda_mqgdo}) for the formulation of data- and equation-assisted schemes (see e.g. \cite{Majda2014, Qi2015}) for filtering and prediction. This step should be important for the short-term prediction \cite{cousins_sapsis, cousinsSapsis2015_JFM} and probabilistic quantification  \cite{mohamad2015, mohamad2016a, mohamad2016b} of systems undergoing extreme responses.

\subsection*{Acknowledgments}
The authors have been supported through the National Science Foundation (NSF EAGER ECCS 15-1462254), the Air Force Office of Scientific Research (AFOSR YIP 16RT0548), and the Office of Naval Research (ONR YIP N00014-15-1-2381). TPS thanks Prof. Kevrekidis, Prof. Giannakis, and Prof. Harlim for numerous stimulating discussions and comments. We also thank Dr. Franzke for providing the code for the climate model that was used to demonstrate the developed methods.  

\bibliography{library}

\begin{thebibliography}{10}

\bibitem{sapsis_majda_mqg}
T.~P. Sapsis and A.~J. Majda.
\newblock {A statistically accurate modified quasilinear Gaussian closure for
  uncertainty quantification in turbulent dynamical systems}.
\newblock {\em Physica D}, 252:34--45, 2013.

\bibitem{sapsis_majda_tur}
T.~P. Sapsis and A.~J. Majda.
\newblock {Statistically Accurate Low Order Models for Uncertainty
  Quantification in Turbulent}.
\newblock {\em Proceedings of the National Academy of Sciences},
  110:13705--13710, 2013.

\bibitem{sapsis_majda_qgdo}
T.~P. Sapsis and A.~J. Majda.
\newblock {Blended reduced subspace algorithms for uncertainty quantification
  of quadratic systems with a stable mean state}.
\newblock {\em Physica D}, 258:61, 2013.

\bibitem{majda11}
A.~J. Majda.
\newblock {Challenges in Climate Science and Contemporary Applied Mathematics}.
\newblock {\em Communications on Pure and Applied Mathematics}, 65:920, 2012.

\bibitem{Majda_filter}
A.~J. Majda and J.~Harlim.
\newblock {\em {Filtering Complex Turbulent Systems}}.
\newblock Cambridge University Press, 2012.

\bibitem{branic_majda}
M.~Branicki and A.~J. Majda.
\newblock {Quantifying uncertainty for predictions with model error in
  non-Gaussian systems with intermittency}.
\newblock {\em Nonlinearity}, 25:2543, 2012.

\bibitem{Gershgorin201032}
B.~Gershgorin, J.~Harlim, and A.~J. Majda.
\newblock {Improving filtering and prediction of spatially extended turbulent
  systems with model errors through stochastic parameter estimation}.
\newblock {\em Journal of Computational Physics}, 229(1):32--57, 2010.

\bibitem{kalman60}
R.~E. Kalman.
\newblock {A new approach to linear filtering and prediction problems}.
\newblock {\em Trans ASME J. Basic Eng.}, 82:35--45, 1960.

\bibitem{kalman61}
R.~E. Kalman and R.~S. Bucy.
\newblock {New results in filtering and prediction theory}.
\newblock {\em Trans ASME J. Basic Eng.}, 83:95--108, 1961.

\bibitem{kitagawa84}
G.~Kitagawa and W.~Gersch.
\newblock {A smoothness priors modeling of time series with trend and
  seasonality}.
\newblock {\em J. Am. Stat. Assoc.}, 79:378--389, 1984.

\bibitem{jones84}
R.~H. Jones.
\newblock {Fitting multivariate models to unequally spaced data}.
\newblock In {\em Time Series Analysis of Irregularly Observed Data}.
  Springer-Verlag, New York, 1984.

\bibitem{shumway88}
R.~H. Shumway.
\newblock {\em {Applied statistical time series analysis}}.
\newblock Prentice-Hall, 1988.

\bibitem{Bongard2007}
J~Bongard and H~Lipson.
\newblock {Automated reverse engineering of nonlinear dynamical systems}.
\newblock {\em Proceedings of the National Academy of Sciences of the United
  States of America}, 104(24):9943--9948, 2007.

\bibitem{Schmidt09}
Michael Schmidt and Hod Lipson.
\newblock {Distilling Free-Form Natural Laws from Experimental Data}.
\newblock {\em Science}, 324(5923), 2009.

\bibitem{brunton16}
Steven~L Brunton, Joshua~L Proctor, and J~Nathan Kutz.
\newblock {Discovering governing equations from data by sparse identification
  of nonlinear dynamical systems.}
\newblock {\em Proceedings of the National Academy of Sciences of the United
  States of America}, 113(15):3932--7, apr 2016.

\bibitem{Majda2013}
Andrew~J Majda and John Harlim.
\newblock {Physics constrained nonlinear regression models for time series}.
\newblock {\em Nonlinearity}, 26(1):201--217, jan 2013.

\bibitem{Peavoy15}
Daniel Peavoy, Christian~L.E. Franzke, and Gareth~O. Roberts.
\newblock {Systematic physics constrained parameter estimation of stochastic
  differential equations}.
\newblock {\em Computational Statistics {\&} Data Analysis}, 83:182--199, 2015.

\bibitem{sapsis_majda_mqgdo}
T.~P. Sapsis and A.~J. Majda.
\newblock {Blending Modified Gaussian Closure and Non-Gaussian Reduced Subspace
  methods for Turbulent Dynamical Systems}.
\newblock {\em Journal of Nonlinear Science}, 23:1039, 2013.

\bibitem{lorenz69}
E.~N. Lorenz.
\newblock {Atmospheric predictability as revealed by naturally occurring
  analogues}.
\newblock {\em J. Atmos. Sci.}, 26:636--646, 1969.

\bibitem{zoth89}
Z.~Toth.
\newblock {Long-range weather forecasting using an analog approach}.
\newblock {\em J. Climate}, 2:594--607, 1989.

\bibitem{sugihara90}
G.~Sugihara and R.~M. May.
\newblock {Nonlinear forecasting as a way of distinguishing chaos from
  measurement error in time series.}
\newblock {\em Nature}, 344(6268):734--741, 1990.

\bibitem{xavier07}
P.~K. Xavier and B.~N. Goswami.
\newblock {An analog method for real-time forecasting of summer monsoon
  subseasonal variability}.
\newblock {\em Mon. Wea. Rev.}, 135:4149--4160, 2007.

\bibitem{Berry2015}
T.~Berry, D.~Giannakis, and J.~Harlim.
\newblock {Nonparametric forecasting of low-dimensional dynamical systems}.
\newblock {\em Physical Review E}, 91(3):032915, mar 2015.

\bibitem{Berry2015b}
T.~Berry and J.~Harlim.
\newblock {Nonparametric Uncertainty Quantification for Stochastic Gradient
  Flows}.
\newblock {\em SIAM/ASA Journal on Uncertainty Quantification}, 3(1):484--508,
  jun 2015.

\bibitem{Berry2016}
Tyrus Berry and John Harlim.
\newblock {Forecasting Turbulent Modes with Nonparametric Diffusion Models:
  Learning from Noisy Data}.
\newblock {\em Physica D}, 320:57--76, 2016.

\bibitem{coifman05a}
R~R Coifman, S~Lafon, A~B Lee, M~Maggioni, B~Nadler, F~Warner, and S~W Zucker.
\newblock {Geometric diffusions as a tool for harmonic analysis and structure
  definition of data: diffusion maps.}
\newblock {\em Proceedings of the National Academy of Sciences of the United
  States of America}, 102(21):7426--31, may 2005.

\bibitem{coifman06}
R.~R. Coifman and S.~Lafon.
\newblock {Geometric harmonics: A novel tool for multiscale out-of-sample
  extension of empirical functions}.
\newblock {\em Applied and Computational Harmonic Analysis}, 21(1):31--52, jul
  2006.

\bibitem{Sonday2010}
B.~E. Sonday, A.~Singer, C.~W. Gear, and I.~G. Kevrekidis.
\newblock {Manifold learning techniques and model reduction applied to
  dissipative pdes}.
\newblock {\em Unpublished}, pages 1--20, 2010.

\bibitem{Chiavazzo2014}
E.~Chiavazzo, C.~Gear, C.~Dsilva, N.~Rabin, and I.~Kevrekidis.
\newblock {Reduced Models in Chemical Kinetics via Nonlinear Data-Mining}.
\newblock {\em Processes}, 2(1):112--140, 2014.

\bibitem{zhao15}
Z.~Zhao and D.~Giannakis.
\newblock {Analog forecasting with dynamics-adapted kernels}.
\newblock {\em Nonlinearity}, page Submitted, 2015.

\bibitem{rasmu05}
C.~E. Rasmussen and C.~K.~I. Williams.
\newblock {\em {Gaussian processes in machine learning}}.
\newblock The MIT Press, Cambridge, MA, 2005.

\bibitem{forrester08}
A.~Forrester, A.~Sobester, and A.~Keane.
\newblock {\em {Engineering design via surrogate modelling: a practical
  guide}}.
\newblock John Wiley and Sons, New York, NY, 2008.

\bibitem{perdikarisPRSA2015}
P.~Perdikaris, D.~Venturi, J.~O. Royset, and G.~E. Karniadakis.
\newblock {Multi-fidelity modelling via recursive co-kriging and
  Gaussian-Markov random fields.}
\newblock {\em Proceedings. Mathematical, physical, and engineering sciences /
  the Royal Society}, 471(2179):20150018, jul 2015.

\bibitem{bilionis12}
I.~Bilionis and N.~Zabaras.
\newblock {Multi-output local Gaussian process regression: Applications to
  uncertainty quantification}.
\newblock {\em J. Comput. Phys.}, 231:5718--5746, 2012.

\bibitem{PChen15}
P.~Chen, N.~Zabaras, and I.~Bilionis.
\newblock {Uncertainty propagation using infinite mixture of Gaussian processes
  and variational Bayesian inference}.
\newblock {\em J. Comput. Phys.}, 284:291--333, 2015.

\bibitem{murphy12}
K.~P. Murphy.
\newblock {\em {Machine Learning: A Probabilistic Perspective}}.
\newblock The MIT Press, 2012.

\bibitem{Kuramoto1976}
Y.~Kuramoto and T.~Tsuzuki.
\newblock {Persistent Propagation of Concentration Waves in Dissipative Media
  Far from Thermal Equilibrium}.
\newblock {\em Progress of Theoretical Physics}, 55(2):356--369, feb 1976.

\bibitem{Sivashinsky1977}
G.I. Sivashinsky.
\newblock {Nonlinear analysis of hydrodynamic instability in laminar
  flames—I. Derivation of basic equations}.
\newblock {\em Acta Astronautica}, 4(11-12):1177--1206, nov 1977.

\bibitem{blonigan14}
P.~J. Blonigan and Q.~Wang.
\newblock {Least Squares Shadowing Sensitivity Analysis of a Modified
  Kuramoto-Sivashinsky Equation}.
\newblock {\em Chaos Sol. Fract.}, 2014.

\bibitem{Kevrekidis90}
I.~G. Kevrekidis, B.~Nicolaenko, and J.~C. Scovel.
\newblock {Back in the saddle again: a computer assisted study of the
  Kuramoto-Sivashinsky equation}.
\newblock {\em SIAM J. Appl. Math.}, 50(3):760--790, 1990.

\bibitem{Lorenz96}
E.~N. Lorenz.
\newblock {Predictability - a problem partly solved}.
\newblock In {\em Proceedings on Predictability}, pages 1--18. ECMWF, sep 1996.

\bibitem{majda_info}
A.~J. Majda, R.~V. Abramov, and M.~J. Grote.
\newblock {\em {Information Theory and Stochastics for Multiscale Nonlinear
  Systems}}, volume~25 of {\em CRM Monograph Series}.
\newblock American Mathematical Society, 2005.

\bibitem{Crommelin2004}
D.~T. Crommelin and a.~J. Majda.
\newblock {Strategies for Model Reduction: Comparing Different Optimal Bases}.
\newblock {\em Journal of the Atmospheric Sciences}, 61(17):2206--2217, 2004.

\bibitem{fran_majd}
C.~Franzke, A.~J. Majda, and E.~Vanden-Eijnden.
\newblock {Low-order Stochastic Mode Reduction for a Realistic Barotropic Model
  Climate}.
\newblock {\em J. Atmos. Sci.}, 62:1722, 2005.

\bibitem{selten95}
F.~M. Selten.
\newblock {An efficient description of the dynamicsl of barotropic flow}.
\newblock {\em J. Atmos. Sci.}, 52:915, 1995.

\bibitem{thom98}
D.~W.~J. Thompson and J.~M. Wallace.
\newblock {The Arctic Oscillation signature in the wintertime geopotential
  height and temperature fields}.
\newblock {\em Geophysical Research Letters}, 25:1297, 1998.

\bibitem{Barn87}
A.~G. Barnston and R.~E. Livezey.
\newblock {Classification, Seasonality and Persistence of Low-Frequency
  Atmospheric Circulation Patterns}.
\newblock {\em Monthly Weather Review}, 115:1083, 1987.

\bibitem{wall81}
J.~M. Wallace and D.~S. Gutzler.
\newblock {Teleconnections in the Geopotential Height Field during the Northern
  Hemisphere Winter}.
\newblock {\em Monthly Weather Review}, 109:784, 1981.

\bibitem{mo86}
K.~C. Mo and R.~E. Livezey.
\newblock {Tropical-extratropical geopotential height teleconnections during
  the Northern Hemisphere winter}.
\newblock {\em Monthly Weather Review}, 114:2488, 1986.

\bibitem{rasmusparse05}
C.~E. Rasmussen and J.~Qui{\~{n}}onero-Candela.
\newblock {A unifying view of sparse approximate Gaussian process regression}.
\newblock {\em J. Mach. Learn. Rsrch}, 6:1939--1959, 2005.

\bibitem{cressie93}
N.~A.~C. Cressie.
\newblock {\em {Statistics for Spatial Data}}.
\newblock Wiley, New York, 1993.

\bibitem{sapsis11a}
T.~P. Sapsis.
\newblock {Attractor local dimensionality, nonlinear energy transfers, and
  finite-time instabilities in unstable dynamical systems with applications to
  2D fluid flows}.
\newblock {\em Proceedings of the Royal Society A}, 469(2153):20120550, 2013.

\bibitem{Majda2014}
A.~J. Majda, D.~Qi, and T.~P. Sapsis.
\newblock {Blended particle filters for large-dimensional chaotic dynamical
  systems}.
\newblock {\em Proceedings of the National Academy of Sciences},
  111(21):7511--7516, 2014.

\bibitem{Qi2015}
D.~Qi and A.~J. Majda.
\newblock {Blended particle methods with adaptive subspaces for filtering
  turbulent dynamical systems}.
\newblock {\em Physica D: Nonlinear Phenomena}, 299:21--41, 2015.

\bibitem{cousins_sapsis}
W.~Cousins and T.~P. Sapsis.
\newblock {Quantification and prediction of extreme events in a one-dimensional
  nonlinear dispersive wave model}.
\newblock {\em Physica D}, 280:48--58, 2014.

\bibitem{cousinsSapsis2015_JFM}
W.~Cousins and T.~P. Sapsis.
\newblock {Reduced order precursors of rare events in unidirectional nonlinear
  water waves}.
\newblock {\em Journal of Fluid Mechanics}, 790:368--388, 2016.

\bibitem{mohamad2015}
M.~A. Mohamad and T.~P. Sapsis.
\newblock {Probabilistic description of extreme events in intermittently
  unstable systems excited by correlated stochastic processes}.
\newblock {\em SIAM ASA J. of Uncertainty Quantification}, 3:709--736, 2015.

\bibitem{mohamad2016a}
M.~A. Mohamad and T.~P. Sapsis.
\newblock {Probabilistic response and rare events in Mathieu's equation under
  correlated parametric excitation}.
\newblock {\em Ocean Engineering Journal}, 120:289--297, 2016.

\bibitem{mohamad2016b}
M.~A. Mohamad, W.~Cousins, and T.~P. Sapsis.
\newblock {A probabilistic decomposition-synthesis method for the
  quantification of rare events due to internal instabilities}.
\newblock {\em Journal of Computational Physics}, 322:288--308, 2016.

\end{thebibliography}
\bibliographystyle{unsrt}

\end{document}